\documentclass[letterpaper,twocolumn,10pt]{article}
\usepackage{usenix-2020-09}

\usepackage[T1]{fontenc}
\usepackage[utf8]{inputenc}
\usepackage{csquotes}
\usepackage{siunitx}
\usepackage{eurosym}
\usepackage{amstext} %
\DeclareRobustCommand{\officialeuro}{%
	\ifmmode\expandafter\text\fi
	{\fontencoding{U}\fontfamily{eurosym}\selectfont e}}
\usepackage{adjustbox}
\usepackage{nicefrac}
\usepackage[toc,xindy,postdot,automake]{glossaries-extra}
\setglossarystyle{listdotted}

\newglossarystyle{listdottedskip}{%
	\setglossarystyle{list}%
	\renewcommand*{\glossaryentryfield}[5]{%
		\item[]\makebox[\glslistdottedwidth][l]{\emph{\glstarget{##1}{##2}}%
			\unskip\leaders\hbox to 2.9mm{\hss.}\hfill\strut}\parbox[t]{\textwidth-\glslistdottedwidth-\labelsep}{##3}}%
	\renewcommand*{\glossarysubentryfield}[6]{%
		\item[]\makebox[\glslistdottedwidth][l]{\glstarget{##2}{##3}%
			\unskip\leaders\hbox to 2.9mm{\hss.}\hfill\strut}##4}%
}

\setglossarystyle{listdottedskip}
\makeglossaries

\setabbreviationstyle[acronym]{long-short-user}

\ifx\noglossaryreferences\undefined

\renewcommand*{\glsxtruserparen}[2]{%
	\glsxtrfullsep{#2}%
	\glsxtrparen
	{#1\ifglshasfield{\glsxtruserfield}{#2}{;
			\expandafter\citep\expandafter{\glscurrentfieldvalue}}{}}%
}

\glsdefpostdesc{acronym}{%
	\ifglshasfield{\glsxtruserfield}{\glscurrententrylabel}%
	{~\expandafter\citep\expandafter{\glscurrentfieldvalue}}%
	{}%
}

\else

\renewcommand*{\glsxtruserparen}[2]{%
	\glsxtrfullsep{#2}\glsxtrparen{#1}}

\fi

\makeatletter
\newcommand*{\glsplainhyperlink}[2]{%
	\colorlet{currenttext}{.}%
	\colorlet{currentlink}{\@linkcolor}%
	\hypersetup{linkcolor=currenttext}%
	\hyperlink{#1}{#2}%
	\hypersetup{linkcolor=currentlink}%
}
\let\@glslink\glsplainhyperlink

\makeatother

\newacronym{dos}{DoS}{denial-of-service}
\newacronym{tcp}{TCP}{Transmission Control Protocol}
\newacronym{lru}{LRU}{least-recently used}
\newacronym{www}{WWW}{World Wide Web}
\newacronym{dag}{DAG}{directed acyclic graph}
\newacronym{ipfs}{IPFS}{InterPlanetary File System}
\newacronym{saas}{SaaS}{Software as a Service}
\newacronym{paas}{PaaS}{Platform as a Service}
\newacronym{aws}{AWS}{Amazon Web Services}
\newacronym[user1={OASISCommitteeDraft2008}]{saml}{SAML}{Security Assertion Markup Language}
\newacronym{idp}{IdP}{Identity Provider}
\newacronym{idm}{IdM}{Identity Management}
\newacronym[user1={Housley2002}]{crl}{CRL}{Certificate Revocation List}
\newacronym[user1={Myers1999}]{ocsp}{OCSP}{Online Certificate Status Protocol}
\newacronym{sso}{SSO}{Single Sign-on}
\newacronym{sp}{SP}{Service Provider}
\newacronym{stork}{STORK}{Secure idenTity acrOss boRders linKed}
\newacronym{pki}{PKI}{Public Key Infrastructure}
\newacronym{soap}{SOAP}{Simple Object Access Protocol}
\newacronym{iam}{IAM}{Identity and Access Management}
\newacronym{uma}{UMA}{User-Managed Access}
\newacronym{fim}{FIM}{Federated Identity Management}
\newacronym{mdsso}{MDSSO}{Multi-Domain SSO}
\newacronym{wsdl}{WSDL}{Web Services Description Language}
\newacronym{w3c}{W3C}{World Wide Web Consortium}
\newacronym{oasis}{OASIS}{Organization for the Advancement of Structured Information Standards}
\newacronym[user1={Hardt2012}]{oauth}{OAuth}{Open Authorization}
\newacronym[user1={OASISStandard2013}]{xacml}{XACML}{eXtensible Access Control Markup Language}
\newacronym{ldap}{LDAP}{Lightweight Directory Access Protocol}
\newacronym{tgt}{TGT}{Ticket Granting Ticket}
\newacronym{pep}{PEP}{Policy Enforcement Point}
\newacronym{pip}{PIP}{Policy Information Point}
\newacronym{pdp}{PDP}{Policy Decision Point}
\newacronym{pap}{PAP}{Policy Administration Point}
\newacronym{prp}{PRP}{Policy Retrieval Point}
\newacronym{ppl}{PPL}{PrimeLife Policy Language}
\newacronym{it}{IT}{Information Technology}
\newacronym{ibac}{IBAC}{identity-based Access control}
\newacronym{rbac}{RBAC}{role-based access control}
\newacronym{abac}{ABAC}{attribute-based access control}
\newacronym{epal}{EPAL}{Enterprise Privacy Authorization Language}
\newacronym{abe}{ABE}{Attribute-Based Encryption}
\newacronym{fpe}{FPE}{Format-Preserving Encryption}
\newacronym[user1=Bethencourt2007]{cp-abe}{CP-ABE}{Ciphertext-Policy Attribute-Based Encryption}
\newacronym[user1=Goyal2006]{kp-abe}{KP-ABE}{Key-Policy Attribute-Based Encryption}
\newacronym{iot}{IoT}{Internet of Things}
\newacronym{ibe}{IBE}{Identity-Based Encryption}
\newacronym{idemix}{idemix}{IBM Identity Mixer}
\newacronym{acl}{ACL}{Access Control List}
\newacronym[user1={TheSUNFISHConsortium2017}]{sunfish}{SUNFISH}{SecUre iNFormatIon SHaring in federated heterogeneous private clouds}
\newacronym{secpal}{SecPAL}{Security Policy Assertion Language}
\newacronym{tc}{TC}{Trusted Computing}
\newacronym{pii}{PII}{Personally Identifiable Information}
\newacronym{toe}{TOE}{Target of Evaluation}
\newacronym{ppa}{PPA}{Privacy-Preserving Authentication}
\newacronym{sep}{SEP}{Sign-Encrypt-Prove}
\newacronym{cpa}{CPA}{Chosen-Plaintext Attack}
\newacronym{srp}{SRP}{Sign-Randomize-Proof}
\newacronym{cca2}{CCA2}{Adaptive Chosen-Ciphertext Attack}
\newacronym{2fa}{2FA}{Two-Factor Authentication}
\newacronym{pow}{PoW}{Proof-of-Work}
\newacronym{rest}{REST}{REpresentational State Transfer}
\newacronym{mitm}{MITM}{Man-In-The-Middle}
\newacronym{ca}{CA}{Certificate Authority}
\newacronym{vpn}{VPN}{Virtual Private Network}
\newacronym{tee}{TEE}{Trusted Execution Environment}
\newacronym{tpm}{TPM}{Trusted Platform Module}
\newacronym{apk}{APK}{Android Application Package}
\newacronym{jni}{JNI}{Java Native Interface}
\newacronym{art}{ART}{Android Runtime}
\newacronym{icc}{ICC}{Inter-Component Communication}
\newacronym{os}{OS}{Operating System}
\newacronym{lrs}{LRS}{Linkable Ring Signature}
\newacronym{trs}{TRS}{Traceable Ring Signature}
\newacronym{drs}{DRS}{Deniable Ring Signature}
\newacronym{zkip}{ZKIP}{Zero-Knowledge Interactive Proof}
\newacronym{p2p}{P2P}{peer-to-peer}
\newacronym{nat}{NAT}{network address translator}
\newacronym{dht}{DHT}{distributed hash table}
\newacronym{zk}{ZK}{Zero-Knowledge}
\newacronym{pabc}{PABC}{Privacy-Preserving Attribute-Based Credential}
\newacronym{nizk}{NIZK}{Non-Interactive Zero-Knowledge}
\newacronym{pcp}{PCP}{Probabilistically Checkable Proof}
\newacronym{fhe}{FHE}{Fully-Homomorphic Encryption}
\newacronym{qap}{QAP}{Quadratic Arithmetic Programs}
\newacronym{cpc}{CPC}{Credential Pseudonymous Certificate}
\newacronym{xml}{XML}{Extensible Markup Language}
\newacronym{daa}{DAA}{Direct Anonymous Attestation}
\newacronym{atn}{ATN}{Automated Trust Negotiation}
\newacronym{lsss}{LSSS}{Linear Secret Sharing Scheme}
\newacronym{ecc}{ECC}{Elliptic Curve Cryptography}
\newacronym{dac}{DAC}{Discretionary Access Control}
\newacronym{geoxacml}{GeoXACML}{Geospatial eXtensible Access Control Markup Language}
\newacronym{abam}{ABAM}{Attribute-Based Access Matrix}
\newacronym{tp-graph}{TP-Graph}{Token-Provisioning Graph}
\newacronym{mac}{MAC}{Mandatory Access Control}
\newacronym{gm}{GM}{Group Manager}
\newacronym{pm}{PM}{Policy Machine}
\newacronym{hgabac}{HGABAC}{Hierarchical Group and Attribute-Based Access Control}
\newacronym[\glslongpluralkey={Trusted Third Parties}]{ttp}{TTP}{Trusted Third Party}
\newacronym{ktaa}{k-TAA}{k-Times Anonymous Authentication}
\newacronym{ip}{IP}{Internet Protocol}
\newacronym{rl}{RL}{Revocation List}
\newacronym{ve}{VE}{Verifiable Encryption}
\newacronym{vlr}{VLR}{Verifier-Local Revocation}
\newacronym{aip}{AIP}{Accountable Internet Protocol}
\newacronym{acl-credential}{ACL}{Anonymous Credentials Light}
\newacronym{crs}{CRS}{Common Reference String}
\newacronym{grbac}{GRBAC}{Generalized Role-Based Access Control}
\newacronym{trbac}{TRBAC}{Temporal Role-Based Access Control}
\newacronym{gtrbac}{GTRBAC}{Generalized Temporal Role-Based Access Control}
\newacronym{sod}{SoD}{Separation of Duty}
\newacronym{arbac}{ARBAC}{Administrative Role Based Access Control}
\newacronym{blp}{BLP}{Bell-LaPadula}
\newacronym{iiot}{IIoT}{Industrial Internet of Things}
\newacronym{mls}{MLS}{Multilevel security}
\newacronym{mic}{MIC}{Mandatory Integrity Control}
\newacronym{ncsc}{NCSC}{National Computer Security Association}
\newacronym{sql}{SQL}{Structured Query Language}
\newacronym{spbac}{SPBAC}{Security Property Based Administrative Controls}
\newacronym[user1={eu:gdpr}]{gdpr}{GDPR}{General Data Protection Regulation}
\newacronym{orbac}{OrBAC}{Organization-Based Access Control}
\newacronym{o2o}{O2O}{Organisation to Organisation}
\newacronym{vpo}{VPO}{Virtual Private Organisation}
\newacronym{rsso}{RSSO}{Role Single-Sign On}
\newacronym{api}{API}{Application Programming Interface}
\newacronym{cca}{CCA}{Chosen Ciphertext Attack}
\newacronym{hve}{HVE}{Hidden Vector Encryption}
\newacronym{cps}{CPS}{Cyber-Physical System}
\newacronym{gid}{GID}{Global Identifier}
\newacronym{aabe}{AABE}{Anonymous Attribute-Based Encryption}
\newacronym{aabpre}{AABPRE}{Anonymous Attribute-Based Proxy Re-encryption}
\newacronym{cp-abpre}{CP-ABPRE}{Ciphertext-Policy Attribute-Based Proxy Re-encryption}
\newacronym{cdn}{CDN}{Content Distribution Networks}
\newacronym[\glslongpluralkey={Attribute Authorities}]{aa}{AA}{Attribute Authority}
\newacronym{dbdh}{DBDH}{Decisional Bilinear Diffie-Hellman}
\newacronym{pbdhe}{PBDHE}{Parallel Bilinear Diffie-Hellman Exponent}
\newacronym{scada}{SCADA}{Supervisory Control and Data Acquisition}
\newacronym{ics}{ICS}{Industrial Control System}
\newacronym{cms}{CMS}{Cyber Manufacturing System}
\newacronym{wsn}{WSN}{Wireless Sensor Network}
\newacronym{iaik}{IAIK}{Institute for Applied Information Processing and Communications}
\newacronym{oabe}{OABE}{Outsourced Attribute-Based Encryption}
\newacronym{ka}{KA}{Key Authority}
\newacronym{rs}{RS}{Re-Encryption Server}
\newacronym{ds}{DS}{Decryption Server}
\newacronym{cl}{CL}{Client}
\newacronym{do}{DO}{Data Owner}
\newacronym{das}{DaS}{Data Source}
\newacronym{sts}{StS}{Storage Server}
\newacronym{ec}{EC}{Elliptic Curve}
\newacronym{mpc}{MPC}{multi-party computation}
\newacronym{faas}{FaaS}{Federation as a Service}
\newacronym{abc}{ABC}{Attribute-Based Credential}
\newacronym{fe}{FE}{Functional Encryption}
\newacronym{arl}{ARL}{Attribute Revocation List}
\newacronym{rp}{RP}{Resource Provider}
\newacronym{aes}{AES}{Advanced Encryption Standard}
\newacronym{ss}{SS}{Storage Server}
\newacronym{iv}{IV}{Initialisation Vector}
\newacronym{soc}{SoC}{System on Chip}
\newacronym[user1={RFC7519}]{jwt}{JWT}{JSON Web Token}
\newacronym{ecdsa}{ECDSA}{Elliptic Curve Digital Signature Algorithm}
\newacronym{dnf}{DNF}{Disjunctive Normal Form}
\newacronym{http}{HTTP}{Hypertext Transfer Protocol}
\newacronym{ico}{ICO}{initial coin offering}
\newacronym{id}{ID}{identifier}
\newacronym{sfs}{SFS}{Self-certifying File System}
\newacronym{xor}{XOR}{exclusive OR}

\usepackage{subfig}
\usepackage{microtype}
\microtypesetup{activate={true,nocompatibility},final,tracking=true,kerning=false,spacing=false,factor=1100,stretch=10,shrink=70}
\usepackage[para]{footmisc}
\usepackage[style=numeric,mincitenames=1,backend=biber,natbib=true]{biblatex}
\begin{document}
\title{Total Eclipse of the Heart -- Disrupting the InterPlanetary File System} %
\date{}

\author{
	{\rm Bernd Pr\"unster}\\
	A-SIT Secure Information\\Technology Center Austria
	\and
	{\rm Alexander Marsalek}\\
	Graz University of Technology
	 \and
	 {\rm Thomas Zefferer}\\
	A-SIT Plus GmbH
} %

\maketitle

\definecolor{rt_attackers}{RGB}{255,0,0}
\definecolor{rt_others}{RGB}{0,255,0}
\definecolor{swarm_attackers}{RGB}{255,128,0}
\definecolor{swarm_others}{RGB}{0,0,255}
\definecolor{bucket_attackers}{RGB}{255,0,0}
\definecolor{bucket_others}{RGB}{0,255,0}
\definecolor{bucket_empty}{RGB}{204,204,204}
\begin{abstract}
	Peer-to-peer networks are an attractive alternative to classical client-server architectures in several fields of application such as voice-over-IP telephony and file sharing.
	Recently, a new peer-to-peer solution called the \gls{ipfs} has attracted attention, which promises to re-decentralise the Web.
	Being increasingly used as a stand-alone application, \gls{ipfs} has also emerged as the technical backbone of various other decentralised solutions and was even used to evade censorship.
	Decentralised applications serving millions of users rely on \gls{ipfs} as one of their crucial building blocks.
	This popularity makes IPFS attractive for large-scale attacks. %
	We have identified a conceptual issue in one of \gls{ipfs}'s core libraries and demonstrate their exploitation by means of a successful end-to-end attack.
	We evaluated this attack against the IPFS reference implementation on the public IPFS network, which is used by the average user to share and consume IPFS content.
	Results obtained from mounting this attack on live \gls{ipfs} nodes show that  arbitrary \gls{ipfs} nodes can be eclipsed, i.e. isolated from the network, with moderate effort and limited resources.
	Compared to similar works, 
	we show that our attack scales linearly even beyond current network sizes and can disrupt the entire public \gls{ipfs} network with alarmingly low effort.
	The vulnerability set described in this paper has been assigned CVE-2020-10937\footnote{\url{http://cve.mitre.org/cgi-bin/cvename.cgi?name=2020-10937}}. Responsible disclosure procedures are currently being carried out and have led to mitigations being deployed, with additional fixes to be rolled out in future releases. Public disclosure has already been coordinated.	
\end{abstract}

\section{Introduction}
\glsresetall

Modern computer networks typically rely on one of two fundamental architectural models. 
The client-server model, which is the predominating model in the \gls{www}, clearly distinguishes network nodes into content providers (i.e.~servers) and content consumers (i.e.~clients). 
In fields of application, where a strict separation of roles is undesirable, computer networks based on the \gls{p2p} model have gained ground. 
Entities participating in \gls{p2p} networks are equal to a large extent, enabling decentralised applications.
This, in turn, makes it possible to escape centralised control and governance as illustrated by cryptocurrencies like \emph{Bitcoin}~\citep{nakamotoBitcoinPeertoPeerElectronic2008}, and systems like \emph{Ethereum}\footnote{\url{https://ethereum.org/en/}}, for example.

Recently, a new \gls{p2p}-based solution called the \gls{ipfs} has attracted attention. 
\gls{ipfs} defines itself as a \enquote{peer-to-peer hypermedia protocol designed to make the web faster, safer, and more open}\footnote{\url{https://ipfs.io}}. 
Developed by \emph{Protocol Labs}\footnote{\url{https://protocol.ai}}, the ambitious goal of \gls{ipfs} is to re-decentralise the \gls{www} in order to relieve it from the drawbacks of classical client-server-based architectures. 
To achieve this goal, \gls{ipfs} replicates and distributes content among participants. 

During the past few years, \gls{ipfs} has increasingly gained traction. 
\emph{Protocol Labs} reported a 30x growth in network size in 2019 and millions of users every week consuming \gls{ipfs} content through their HTTP to \gls{ipfs} gateway~\citep{mackinlayIPFSProjectFocus}. 
The report also mentions hundreds of thousands of users actively participating in the \gls{ipfs} core network and hundreds of individual developers contributing every month to the \gls{ipfs} code base on GitHub.

At the same time, \gls{ipfs} has also established itself as the technical foundation for various other decentralised applications. 
For instance, \gls{ipfs} acts as one of the enabling technologies for \emph{Filecoin}~\citep{protocollabsFilecoinDecentralizedStorage2017}, a cryptocurrency developed by {Protocol Labs}, pitched as a \emph{robust foundation for humanity's information}\footnote{\url{https://filecoin.io}}. 
Filecoin has had one of the largest ever \glspl{ico} to date, raising over \$205Mio~\citep{aloisProtocolLabsFiles2017}.
Amongst others, \gls{ipfs} also serves as the technical foundation of \emph{DTube}\footnote{\url{https://d.tube}}, a decentralised video platform with millions of active daily users.
Services like \emph{Textile}\footnote{\url{https://textile.io/}}, for example, seek to lower the barrier to using IPFS, while  \emph{Pinata}\footnote{\url{https://pinata.cloud/}} offers guaranteed data availability and IPFS-based hosting.
Moreover, the cryptocurrency \emph{Ethereum} will be using \emph{libp2p}, a key component of IPFS as the networking layer for the \emph{Ethereum 2.0} network~~\citep{thelibp2pteamLibp2p2020}.
The growing relevance of \gls{ipfs} is also underpinned by the fact that the Opera web browser has added native \gls{ipfs} support on Android~\citep{ayalaIPFSOperaAndroid} recently. 
Finally, \gls{ipfs} has also been used to evade censorship. 
For instance, in 2017 the Catalan independence movement used \gls{ipfs} after being declared illegal~\citep{hillCatalanReferendumApp}.
This increased popularity leads to the general question of the resilience of ungoverned, open \gls{p2p} systems against attacks inherent to this model.
The works of \citet{heilmanEclipseAttacksBitcoin2015},  \citet{marcusLowResourceEclipseAttacks} and \citet{henningsenEclipsingEthereumPeers2019} aiming at Bitcoin and Ethereum nodes clearly show that attack vectors inherent to \gls{p2p} systems can be exploited in practice to isolate individual participants, which is most relevant for cryptocurrency-specific attacks.
Based on the raising popularity of IPFS, these observations lead to the following research question: \emph{Is it possible to design a low-cost, global attack on a decentralised \gls{p2p} system that is used in practice, such that it scales well with network size?}

\paragraph{Contribution and Scope}
This paper presents an end-to-end eclipse attack on IPFS, exploiting a conceptual issue in a core component of IPFS that compromises the system's overall security.
We evaluated our attack against the IPFS reference implementation, \emph{go-ipfs}, version 0.4.23 and the decentralised \gls{p2p} network spanned by those nodes.
In particular, our contribution is fourfold:\\[-1.5em]
\begin{description}
	\item[Attack] We introduce an end-to-end eclipse attack on \gls{ipfs}.
	This attack enables an attacker to single out network nodes of their choice, partition, and disrupt the \gls{ipfs} network.
\\[-2em]	\item[Implementation] We describe successful mounting of the proposed attack on live \gls{ipfs} nodes, even those part of critical infrastructure.
\\[-2em]	\item[Evaluation]  We elaborate on the threat potential of the attack, concluding that even modestly powerful attackers can carry out the attack to disrupt the whole public \gls{ipfs} network.
\\[-2em]	\item[Countermeasures]  We have reported our findings to {Protocol Labs}, resulting in countermeasures being rolled out.
	While the specific attack presented here has since been mitigated, the hardening process is still ongoing, highlighting the sustainable impact of this work on systems used in production.
	IPFS 0.5 released in May 2020 already includes a major rewrite of a previously vulnerable core component.
	The 0.6.1 version of IPFS  introduced a bulk of changes that further contribute to attack resiliency and inflate the cost of our attack by several orders of magnitude. The current 0.7 version finally breaks compatibility with older, vulnerable releases.\\[-1.5em]
\end{description}
The issue we uncovered is a conceptual one and thus has an impact beyond IPFS itself.
Actually weaponising this weakness requires specific attack vectors, three of which were discovered to affect the public {DHT-based IPFS network}, which will be referred to as \emph{IPFS DHT} in this paper.
This is the network a user will interact with when using the official desktop or command-line IPFS distributions downloadable from \url{ipfs.io}.
A detailed explanation on this terminology with respect to the attack scope is provided as part of Section~\ref{sec:eval}, that provides details on how the attack was evaluated.

At the time of discovering the vulnerability enabling our attack (April 2020), go-ipfs 0.4.23 was the most current release.
Due to the modularity of IPFS and the wide use of (parts of) it in other projects, our attack's impact beyond the public DHT-based IPFS network needs to be evaluated on a per-project basis, which is beyond the scope of this work\footnote{For example, the main attack vector exploited for attacking the IPFS DHT is not present in Filecoin, according to Protocol Labs.}.
However, as mitigations have been rolled out, other projects already benefited from the fixes resulting from this work.

\paragraph{Paper Outline}
This paper is structured as follows.
Relevant background information is provided in Section \ref{sec:preliminaries} to support an in-depth understanding of our attack and its consequences.
Details on the attack itself are introduced in Section \ref{sec:attack}.
Subsequently, figures obtained from applying our attack on live \gls{ipfs} nodes to evaluate the attack's feasibility are presented  in Section \ref{sec:eval}.
Finally, we discuss potential countermeasures in Section \ref{sec:countermeasures}, introduce related scientific work in Section \ref{sec:related}, and conclude the paper in Section \ref{sec:conclusions}.
Appendixes provide details on the evaluation procedure, information on ethical aspects of our work, as well as scaled-up versions of figures used throughout the main paper for better legibility. %

\section{Preliminaries}
\label{sec:preliminaries}

To ensure a comprehensive understanding of the attack described in this paper, this section provides necessary background information. 
Section \ref{sec:ipfs} thus provides a technical overview on \gls{ipfs}.
In-depth technical details are introduced in Section \ref{sec:libp2p}, which focuses on the library \emph{libp2p} implementing key functionality of \gls{ipfs} and representing the core target of our attack. 
Finally, Section~\ref{sec:attacks} provides background information on known attack vectors for \gls{p2p} networks, which have inspired our attack.%

\subsection{IPFS}
\label{sec:ipfs}

From a technical perspective, \gls{ipfs} is a distributed, content-addressed file system, where data is not identified by name or path, but by its hash. 
\gls{ipfs} stores all data in a decentralised way overlaying the whole network with a Merkle \gls{dag}~\citep{merkleDigitalSignatureBased1987} to create a navigable structure. 
All content stored in the network and every node participating in the network are assigned a unique \glsentrylong{id} from the same flat \gls{id} space. 
Content \gls{id}s are derived directly from the respective data by computing the data's cryptographic hash value. 
Peers, i.e.~nodes participating in the network, generate an asymmetric cryptographic key pair. 
The public key serves as unique \gls{id} for the peer. 
The private key is used by the peer to sign outgoing data in order to provide receiving nodes evidence on its identity. 
Summarising, \gls{ipfs} builds on the concepts of the \glsentrylong{sfs} introduced by \citet{mazieresSelfCertifyingFileSystem2000} and uses public-key cryptography for the self-certification of objects.

Lacking any central authorities, secure and reliable content and peer discovery is a key challenge. 
\gls{ipfs} implements this  functionality based on a \emph{Kademlia} \gls{dht}~\citep{baumgartKademliaPracticableApproach2007}. 
Accordingly, each node maintains its own routing table containing information about neighbouring nodes. 
This information is structured as binary tree containing mappings of node \gls{id} to network (IP) addresses. 
To find a certain node or specific data, the node traverses its own tree for the required \gls{id}. 
If it is able to discover the required \gls{id}, the node can access the associated information using the assigned network address. 
Otherwise, the node asks peers that are closest to the sought-after information. 
As \gls{ipfs} uses \emph{Kademlia}, the closest nodes can be found by means of its \gls{xor} distance. 
This last step, i.e.~finding the closest nodes, can be repeated until the required node or information is found. 
This approach is proven to be efficient, taking only $ \mathcal{O}(\mathrm{log}_2(\textit{network-size})) $ many requests to locate any content or node.
This efficiency comes at the price of limiting the amount of routing information that can be locally stored.
To compensate for this, \gls{ipfs} features a data structure called the \emph{swarm}, which essentially keeps connections beyond the \gls{dht}.
This is used for the main content-distribution functionality of \gls{ipfs}: First, the swarm is asked for data. 
If someone in the swarm is able to provide it, discovering data is a constant-time operation, if not, the \gls{dht} is used\footnote{\ttfamily \textcolor{blue}{https://github.com/ipfs/go-bitswap/blob/master/docs/how- bitswap-works.md}}. 

\gls{ipfs} has been designed to be fully open and decentralised, thus no central authority guards it. 
Consequently, anyone can join the network and identify using a generated asymmetric cryptographic key pair. 
This obvious strength of \gls{ipfs} is also one of its Achilles heels, facilitating the attack proposed in this paper, as presented in Section~\ref{sec:attacks}.

\subsection{libp2p}
\label{sec:libp2p}

The library \emph{libp2p}\footnote{\url{https://libp2p.io}} is a stand-alone project that was originally an integrated part of IPFS, but has since been externalised.
It encompasses a \gls{dht}, transport abstractions and other components required to build decentralised applications. 
In essence, libp2p serves as basis for various solutions that require a \gls{p2p} network.
Moreover, it supports connecting through \glspl{nat} and to some extent also supports browser-based environments using WebSockets~\citep{fetteWebSocketProtocol2011}. 
\gls{ipfs} is one of many solutions that heavily rely on \emph{libp2p} and its \gls{p2p} functionality.

The attack described in this paper employs a set of vulnerabilities in  \emph{libp2p}. 
Accordingly, this is rather an attack on \emph{libp2p} than on \gls{ipfs}. 
However, we also exploit the way \gls{ipfs} interacts with \emph{libp2p} to increase attack efficiency.
Moreover, \gls{ipfs} is the largest public libp2p-based network, which also serves as infrastructure layer for other decentralised services. 
Since we mounted our attack on the \gls{ipfs} reference implementation, we refer to  \gls{ipfs} throughout this paper, although the attack actually targets \emph{libp2p}. 

\paragraph{\glsentryshort{dht}}
As mentioned above, libp2p's \gls{dht} is based on \emph{Kademlia}~\citep{maymounkovKademliaPeertoPeerInformation2002}.
As of April 2020, libp2p mainly used connection-oriented transport protocols like TCP.
Consequently, nodes are only kept in the local routing table as long as an active network link to this node exists.
The \gls{dht}'s binary tree structure allows for a configurable amount of nodes to occupy each leaf of this tree.
Leaves are referred to as \emph{k-buckets} or simply \emph{buckets}.
The bucket-size parameter is called \emph{k} and is set to 20 in IPFS.

Tree branches can be merged and split on-demand in case buckets are not fully occupied or become overfull.
However, the total number of nodes that can be kept in a local routing table is limited according to Eq.~\eqref{eq:rt-size}.
Currently, libp2p uses SHA2-256 as cryptographic hash function to derive node and content identifiers for \gls{dht} routing, which yields a 256-bit \gls{id} space.
Note that Kademlia proposes a least-recently seen eviction strategy, should a bucket become overfull.
When using connection-oriented transport protocols, this is implemented implicitly using transport-layer keep-alive messages.
Thus, newly connecting peers will only replace others, in case those others become unresponsive.

Another key feature provided by the \gls{dht}'s peer discovery functionality, is \emph{bootstrapping}:
In order to initially join the network, the IP address of at least one node already participating in the network needs to be known.
Once connected to one such pre-known node, a newly joining peer queries this \emph{bootstrap} node's routing table for their own ID.
This prompts a response containing the \gls{id}s and IP addresses of other nodes known to the bootstrap node, which are closest to the newly joining peer (from the bootstrap node's point of view).
As of version 0.4.23, \gls{ipfs} comes pre-configured with eight bootstrap nodes run by {Protocol Labs}.

\vspace*{-0.5em}
\begin{equation}
 {RT\: size} = \textrm{bits}({ID\: space})\times{}k
 \label{eq:rt-size}
\end{equation}

\paragraph{Swarm}
\gls{ipfs} raises the requirement to store connection information that exceeds the \gls{dht}'s limited capacity.
For this,  \gls{ipfs} nodes make use of the so-called \emph{swarm}. 
First and foremost, the \emph{swarm} is the set of all currently active connections and thus a superset of the connections stored in the \gls{dht}.
\gls{ipfs} also uses the swarm to speed-up content discovery by initially querying the whole swarm for content, prior to querying the \gls{dht}\footnote{\ttfamily \textcolor{blue}{https://github.com/ipfs/go-bitswap/blob/master/docs/how-bitswap- works.md\#discovery}}.
On its own, the swarm is unbounded, which could lead to resource exhaustion.
To prevent this, a component called the \emph{connection manager} or \emph{ConnMgr} is in place, as described below.

\paragraph{ConnMgr}
\label{sec:intro:connmgr}

The connection manager is provided by libp2p. 
Its main job is to keep only a sensible amount of open connections. 
This ensures that (a) resource exhaustion is prevented and (b) content discovery and the overall \gls{p2p} network flow can operate efficiently.

Currently, libp2p features a single implementation of the connection manager. 
This implementation traverses the set of active connections (i.e. the swarm) once a minute.
In case more than a configurable threshold of connections are open (called the \texttt{highWater} mark), one connection at a time is trimmed, until the second configured threshold (called \texttt{lowWater}) of open connections is reached\footnote{Whether these connections are incoming or outgoing is irrelevant}.
Recently-established connections (within a configurable \emph{grace period}) are exempt from pruning.
Starting with the libp2p version that ships with \gls{ipfs} 0.4.23, these connections do not count towards the total amount of active connections.
This is crucial, since it prevents a cheap attack where an attacker could simply connect \texttt{highWater} many connections within the grace period to have the ConnMgr unconditionally trim all older connections.

The main challenge with this approach is determining which connections to trim. 
For this, an abstract scoring system is in place that is available to all components interacting with \emph{libp2p}. 
Any interacting component is allowed to add a freely-definable tag to any active connection and award points under this tag.
The general idea behind this approach is to keep highly useful connections open.
For instance, connections in the \gls{dht} are awarded points according to Eq.~\ref{eq:dhtscore}.
This effectively means that closer (according to their \gls{xor} distance) nodes are awarded higher scores and are less likely to be disconnected.

\vspace*{-1em}
{\small 
\begin{equation}
score = 5 + \mathrm{commonXORPrefixLen}(remote\:ID, own\: ID)
\label{eq:dhtscore}
\end{equation}
\vspace{-1em}}

Other sources of points include a relaying subsystem part of libp2p, which is used to help nodes behind firewalls connect to the network:
In short, any node can advertise themselves as relay and offer multiplexing other nodes' connections over an already established link between target node and relay. 
Points can also be awarded by the so-called \emph{Bitswap} subsystem, which is a core component implementing the content-distribution strategy employed by \gls{ipfs}~\citep{benetIPFSContentAddressed2014}. 

While the applied scoring system is essential for the connection manager's functionality, it causes a potential vulnerability: 
If ways can be found to artificially inflate the score of connections to a node, this node will less likely be disconnected, even if it behaves maliciously. 
The attack proposed in this paper employs this vulnerability.
As we have discovered, \emph{Bitswap} awards points even when receiving unsolicited data blocks. 
This can easily result in more points than awarded from some of the lower \gls{dht} buckets (i.e.) those containing the farthest-away nodes.
Given the swarm is a superset of the \gls{dht}, 
this can lead to \gls{dht} connections being trimmed in favour of non-\gls{dht} actively advertising data.
In combination with IPFS's inherent susceptibility to Sybil attacks (see Section~\ref{sec:attacks}), a node can be manipulated into eclipsing itself from the honest network.
As we will show, it is possible to execute such a strategy with extremely low cost.
It is important to note that this is a conceptual flaw with the only variable being the actual resources required to mount an attack.

\subsection{Known Attack Strategies}
\label{sec:attacks}

\gls{ipfs} is a fully open and decentralised solution with no central coordinating or regulating authorities. 
These properties make \gls{ipfs} vulnerable to two different attack strategies, which are not specific for \gls{ipfs}, but apply to any \gls{p2p} network with comparable properties. 
The two attack strategies that make use of these vulnerabilities %
have become known as \emph{Sybil} and \emph{eclipse} attack. 

In the \emph{Sybil} attack~\citep{douceurSybilAttack2002a}, a single attacker presents itself to the network as many seemingly independent nodes by generating multiple identities. 
This can subvert any network operation that works under the assumption of interacting with distinct, non-colluding entities, such as the distributed routing protocol itself. 
Due to its decentralised nature and openness, \gls{ipfs} is conceptually vulnerable to Sybil attacks. 

The second prominent strategy to compromise \gls{p2p} systems is the \emph{eclipse} attack~\citep{singhDefendingEclipseAttacks2004}. 
In essence, the attacker manipulates a node's local routing information, such that any request from or to the victim passes through nodes controlled by the attacker.
For a structured \gls{p2p} network, this requires generating identifiers of a specific distance to the chosen victim's identifier. 
Since node identifiers in \gls{ipfs} are hashed prior to computing this distance, large numbers of \gls{id}s need to be generated and tested to obtain \gls{id}s with suitable distances.
While this may seem infeasible, we show that a brute-force approach to this problem actually scales well, enabling global attacks even for large network sizes\footnote{This scales, since only the network size is relevant, not the size of the ID space.} (see Section~\ref{sec:atk:id}).
Once this is done, as many nodes as required to eclipse a victim using these \gls{id}s can be operated by applying a Sybil attack. %

Any open and fully decentralised \gls{p2p} network is vulnerable to Sybil and eclipse attacks on a conceptual level. 
Fortunately, a variety of countermeasures to these generic attack strategies exist in practice, although as of version 0.4.23, IPFS did not include any.

\section{End-to-End Eclipse Attack on IPFS}
\label{sec:attack}
If one node after another can be eclipsed from the rest of the network and the amount of resources required to keep nodes from reconnecting are low, even an average-powered attacker can disrupt \gls{p2p} networks that are as a whole many orders of magnitude more powerful than the attacker.
This work demonstrates precisely this kind of attack against the live IPFS DHT from two distinct angles. %
We show how we can advance from attacking single nodes to partitioning the network with negligible running costs.
Our implemented end-to-end attack is able to automatically poison any node's routing table on the main IPFS network within minutes, regardless of the network's \emph{churn rate}\footnote{The churn rate is defined as the participant turnover in the network, i.e. how fast participants join and leave the network.} and to fully eclipse an average node in less than an hour with $ \approx75\% $ probability (see Section~\ref{sec:eval}).
The only input required to start the attack is a target node's identifier.
Moreover, IPFS uses a configurable, but otherwise static set of nodes for bootstrapping.
Poisoning the bootstrap node's routing tables (with potentially bogus information) is therefore enough to keep any node that carries out the bootstrap routine from ever interfacing with other legitimate nodes\footnote{unless measures beyond the default behaviour have been explicitly set up}.
At this point, partitioning the IPFS DHT becomes possible.%

\begin{figure}
	\centering
	\includegraphics[width=\columnwidth]{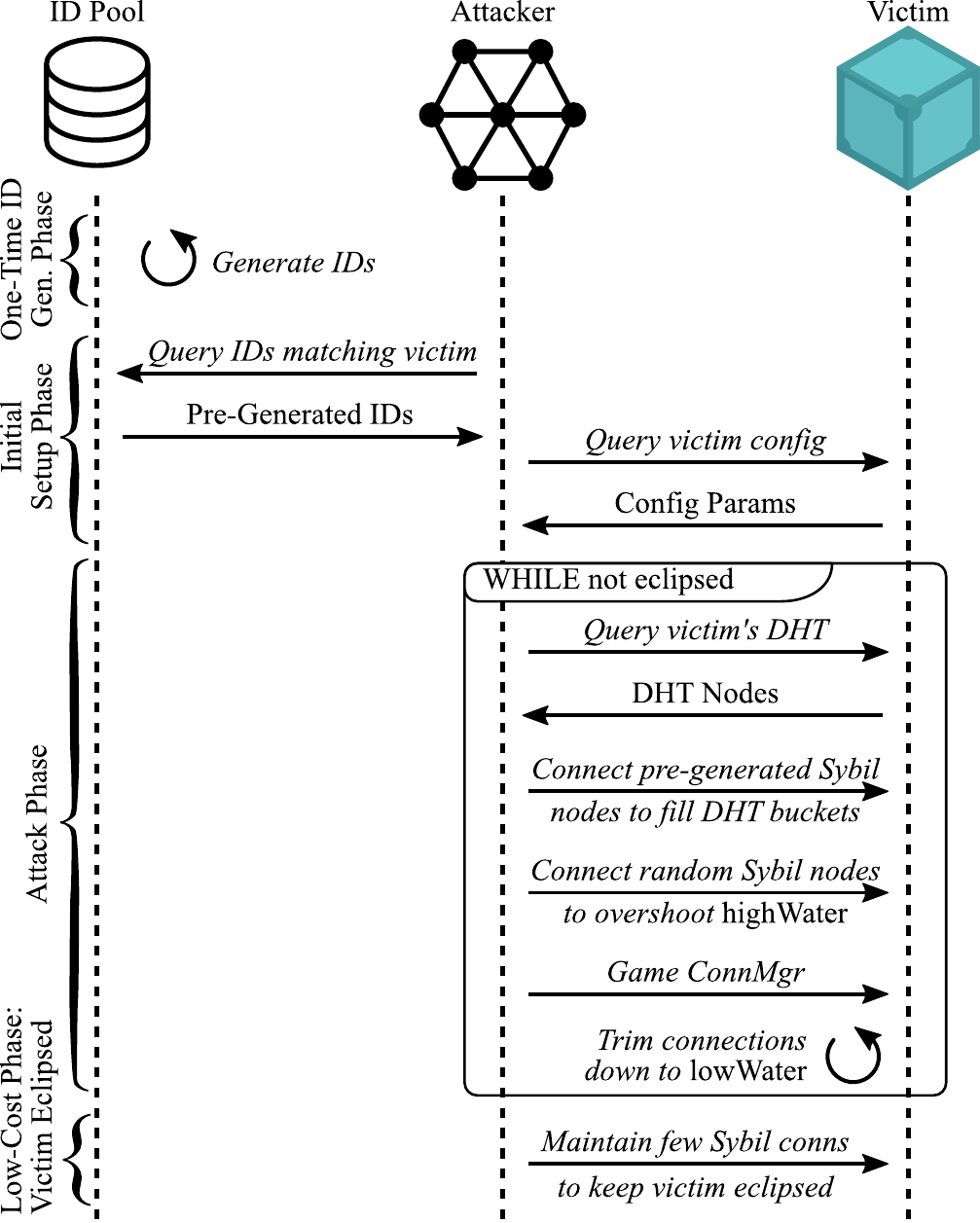}
	\caption{Abstract end-to-end flow of eclipsing an IPFS node}
	\label{fig:seqd}
\end{figure}

Actually weaponising these observations to mount an attack on the live IPFS network consists of three main steps:
	\\[0.2em]	\textbf{ID generation:} In order to eclipse a target's routing table, pre-generated IDs are required.
	Section~\ref{sec:atk:id} describes this aspect in detail, as the cost of ID generation has a direct impact on attack costs.
	\\[0.2em]	\textbf{Probing:} IPFS/libp2p offers a vast amount of configuration options. %
	Thus, we require means to probe a target's state, as detailed in Section~\ref{sec:probing}. %
	\\[0.2em]	\textbf{Gaming the ConnMgr:} The key part of our attack is to trick a target's ConnMgr into trimming all connections to  legitimate nodes, such that only malicious nodes operated by an attacker remain.
	Details on this attack strategy are elaborated in Section~\ref{sec:atk:connmgr}.
	 Once successful, regular nodes can be kept eclipsed with extremely low cost (see Section~\ref{sec:issues:impl}). 
\\[0.4em]	
The actual attack is carried out by a stripped-down libp2p-based node that performs a Sybil attack, which is then used to eclipse the target.
This node requires a set of pre-generated IDs (from the ID generation step), that fit the victim's DHT.
The main attack loop then consists of the following steps:
\begin{enumerate}
	\item \vspace*{-0.4em}Establish as many connections to the target as possible, each with one of the pre-generated identifiers.
	These identifiers cover the lowest 33+ buckets of the victim\footnote{A network of ($ >2^{33} $) of honest nodes would be required to fill this many buckets, while the theoretical maximum is 256 buckets for libp2p.}.
	\item \vspace*{-0.5em} Establish additional connections with randomly generated identifiers to reach a total of \texttt{highWater} many active connections.
	This ensures periodic trimming of connections by the target's ConnMgr.
	\item \vspace*{-0.5em} Messages are sent over each connection to inflate its score, thus tricking the target's ConnMgr into considering these connections more important than those to honest network nodes.
	\item \vspace*{-0.5em} When the target's ConnMgr trims connections to reach \texttt{lowWater} many active connections, only legitimate connections established within the grace period will survive.
	All others will be pruned, leaving mostly those connections established by our malicious libp2p node, since we previously tricked the target's ConnMgr into considering our connections more important than those to honest nodes.
\end{enumerate}
\vspace*{-0.4em}
Since we are able to exploit the ConnMgr to our advantage, our connections will gradually fill up the target's routing table (this takes 2 minutes at most) as well as the swarm since honest ones are pruned. 
When queried for content or other nodes, our attacker nodes will  filter out any information on other (legitimate) nodes from responses to hinder the victim from learning about other nodes.
We rely on continuously probing the target's routing table to receive feedback about the attack's progress (see Section~\ref{sec:probing}).
In addition, we rely on the data obtained during our attack evaluation to draw conclusions about the state of an attack target's swarm (see Section~\ref{sec:eval}).
Once successful, only four connections suffice to keep a regular IPFS node eclipsed, as explained in the following section.
Fig.~\ref{fig:seqd} presents a high-level sequence diagram of the complete end-to-end attack flow, grouping attack steps into higher-level phases. An additional visualisation of the state of a victim node's buckets during normal operation and during the attack is available in Appendix~\ref{app:vis}.

\subsection{Implementation Flaws}
\label{sec:issues:impl}
Apart from conceptual issues %
we have identified the following flaws, that contribute to our attack's performance:
\\[0.2em]
\textbf{Allowing Only Inbound Connections:} Although libp2p differentiates between inbound and outbound links, this has no bearing on the ConnMgr's trimming routine, making it possible to trim all outgoing connections.
\\[0.2em]
\textbf{Unconditional Removal of \gls{dht} Nodes:} Since the ConnMgr does not interpret scores and tags, and \gls{dht} connections do not receive special treatment, it is possible to push all legitimate connections out of the \gls{dht} and replace them with malicious ones.
	This goes against the original Kademlia design, which favours older connections.
	Note that from the \gls{dht}'s point of view, this characteristic is still upheld, but the ConnMgr manually disconnects already known connections that would prevent new ones from entering the \gls{dht}.
\\\textbf{Stateless Connectivity Monitoring:}
	The \gls{dht} is instructed to re-execute a bootstrap manoeuvre to connect to pre-configured bootstrap nodes, whenever less than four open connections remain.
	However, no further action is taken even if this situation becomes stationary.
	This effectively enables an attacker to keep a node eclipsed with as little as four open connections, once an eclipsed state is initially reached. %
	This monitoring loop to check connectivity is executed once a minute.%
\\\textbf{Static Bootstrap Nodes:}
 IPFS relies on a static set of bootstrap nodes.
	Although this set can be configured, a node does not keep track of peers it was once connected to.
	Due to this, a restarting node will always bootstrap against the same set of nodes, regardless of connectivity prior to restarting.
	As a consequence, compromising the default set of bootstrap nodes will affect all nodes as soon as they restart (and not only newly joining nodes).
\\\textbf{Unconditional Trust:}
Although rigorous data integrity checks are a core feature of the main IPFS functionality, the same can not be said with respect to information regarding the network's node.
This meta issue affects many subsystems.
	In short, almost all claims made by a peer about its characteristics and capabilities are taken at face value, even when simple checks could expose cheating.
	As for the \gls{dht}, this makes it easily possible to fill any node's \gls{dht} with bogus IP addresses.

\subsection{ID Generation}
\label{sec:atk:id}
Our attack requires efficient generation of vast amounts of valid identifiers to position nodes at specific distances to \emph{any} node.
Thus, a one-time ID pre-generation routine is run.
libp2p supports RSA and EC keys. 
By generating only EC-based identifiers, both generation times and the amount of storage required to manage a large set of IDs is reduced compared to  RSA. %
To maximize throughput, we 
 increment an integer and interpret it as a private key\footnote{Since identifiers are hashed prior to calculating distances, lack of randomness in the raw key material is not an issue.}.
This approach generates 8-10k keys per second per CPU core based on an \emph{Intel}\textregistered{} \emph{Xeon}\textregistered{} E5-2699 v4 CPU @ 2.20GHz.
Overall throughput is mainly limited by IO speed.%

Our storage format is simple and efficiently searchable: Each generated key is stored to a file named after the first 14 bits of the \gls{dht} identifier corresponding to the key, along with this identifier.
This set of pre-generated identifiers amounts to 29TB of data and encompasses $ \approx 146$Bn individual IDs.
Still, it is possible to efficiently query this database, as the number of entries per file remains manageable.
A dedicated service in charge of producing keys and identifiers corresponding to any node's lowest 33+ \gls{dht} buckets, based on the target node's ID as input.
Answering a query takes less than five minutes\footnote{This could be further sped-up thorugh parallelisation.}. %

\subsection{Probing}
\label{sec:probing}
In order to carry out our attack, ways to continuously probe an attack target's state are required.
Additionally, some static parameters are needed during the attack's setup phase.

\paragraph{Setup Phase}
Naturally, an attack target's IP address is required to connect to it.
This information is obtained by operating a regular IPFS node that is connected to the IPFS network, 
This node is used to query the target's IP address,
 which is then fed into the actual attack carried out on a distinct machine.

Moreover, the bucket size of the target's routing table is required in order to advertise a precisely distributed set of identifiers for routing table poisoning.
Querying the target for any identifier will trigger a response containing $ bucketSize $ many peers, thus providing this information.

While \texttt{lowWater} and \texttt{highWater} marks are theoretically required to perform our attacks, choosing too high values has no impact on attack success rates, but only consumes more resources than necessary.
Given that even critical infrastructure like bootstrap nodes use values of 1000 and 2000, respectively, this does not cause any real issues with respect to attack performance.
However, simply starting with those values and and observing the impact of connection trimming allows for detecting values set too high, which enables reducing each value accordingly.

The interval for the ConnMgr's connection trimming routine is hardcoded to 1 minute.
However, detecting disconnect waves when maintaining hundreds of open connections is trivial,
 as our attack operates a little over \texttt{highWater} many connections to ensure that disconnect waves are triggered.
\\
The grace period used to protect newly established connections is irrelevant for our attack and is thus not probed.

Our attack targets the latest IPFS version (0.4.23) released as of April 27, 2020.
Earlier versions are even easier to eclipse. However, our attack performs a strict superset of the actions required to eclipse earlier versions and thus requires no knowledge of the attack target's IPFS version, except for increased efficiency.
Still, the IPFS protocol defines a message to remotely query a node's version.%

\paragraph{Continuous Probing}
\label{sec:atk:cont-probing}
Our attack requires knowledge about the target's routing table.
In essence, we need to know which buckets are occupied by honest nodes, in order to outperform these nodes from the ConnMgr's point fo view.
As mentioned before, the target will respond with $ bucketSize $ many closest peers to any query for other peers.
We can thus simply traverse the set of pre-generated identifiers used for poisoning the target's routing table and query for one identifier in each bucket.
This way, it is possible to construct a contiguous view of the target's routing table and know precisely which nodes occupy which buckets.
Given that our pre-generated ID set consists of $ \mathbin{\approx}{146\text{Bn}}$  nodes, this easily covers all realistically possible routing table configurations that can ever be encountered.
These queries are performed once during each attack loop.

\subsection{Gaming the ConnMgr}
\label{sec:atk:connmgr}
This section describes the main attack loop and the actions performed to trick a victim's ConnMgr.
The overall goal is to raise the score of the connections made by an attacker above the highest score of any legitimate node connected to the victim.
Based on observation of live IPFS nodes, the score of connections to honest peers will usually range from 0 to around 20.
In order to reach this goal, our attack strategy relies on three sources of points to game the ConnMgr:\\[0.2em]
	\textbf{\gls{dht}:} Each node that occupies a \gls{dht} spot is awarded points according to Eq~\ref{eq:dhtscore}.
	The amount of identifiers we pre-generate is several orders of magnitude larger than the number of nodes participating in the live IPFS \gls{dht}.
	Because of this, simply connecting using these IDs is enough to be assigned a spot in the target's \gls{dht}, as most buckets for those identifiers will be empty.
	We can therefore maintain more than 400 connections\footnote{The remainder of these connections are initially outperformed by honest ones.} that will be awarded enough points to become resident in the target's swarm.
	This, however is significantly less than the required default \texttt{lowWater} value of 600.
	\\\textbf{Bitswap:} As mentioned in Section~\ref{sec:intro:connmgr}, Bitswap awards points for unsolicited content advertisement (which is understandable from a content-distribution perspective).
	We exploit this by continuously advertising an empty block of data.
	This is cheap for an attacker, since sending such a message every few seconds suffices, with no need to process responses and results in 10-16 points.
	Other ways of inflating a connection's score based on Bitswap include re-sending blocks a target previously requested.
	\\\textbf{Relaying:} A virtually unlimited source of ConnMgr points is the relaying subsystem that is used to help nodes located behind \glspl{nat} or firewalls reach the network.
	For one, simply advertising relaying capabilities to the target already awards a fixed amount of two points.
	More importantly, however, actively relaying connections from and to the target awards one point for each relayed connection.
	Given that libp2p supports multiplexing many virtual connections over a single (TCP) link, the number of available ports is not a limiting factor for this strategy.
	Initial experiments have shown that $ >1000 $ connections can be multiplexed over a single link.\\[0.2em]
This last method of obtaining points can be especially devastating, since finding a countermeasure is challenging.
While pathological cases like those from our initial experiments could be detected using heuristics, the general strategy of considering a link supporting many relayed connections important is understandable, especially in real-world settings that include firewalls and \glspl{nat}.
The required resources for creating a relayed connection are minimal:
The only thing that is really required is a (randomly generated) key pair to obtain a valid self-certifying node identifier.
This is then used during the initial handshake when establishing an end-to-end-encrypted connection.
While this comprises computationally somewhat expensive asymmetric cryptographic operations, these have to be performed only once during connection establishment. %
Overhead for the node acting as relay is also moderate.
Given that our attack strategy is based on performing a Sybil attack from a single host %
no actual links between relay and relayed nodes are required, since these nodes are, in fact, virtual.

When combining this way of inflating connection scores with the continuous probing of a target's routing table, a highly efficient attack behaviour can be implemented.

In order to occupy all spots in the target's routing table and, subsequently, eclipse the whole swarm,
any nodes previously connected need to be outperformed.
However, fully poisoning the routing table is prioritised, since the \gls{dht} is used for peer discovery and content routing beyond content distribution.
In order to accomplish this, estimate the highest score of any honest node connected to the target as follows:
\begin{enumerate}
	\item \vspace*{-0.5em}Based on the routing table information, we calculate the highest score over all legitimate peers that reside in the target's routing table according to Eq.~\ref{eq:dhtscore}.
	\item\vspace*{-0.5em} We consider a safety margin of 10 points, meaning that we assume that each honest peer has an additional 10 points awarded from other subsystems, such as Bitswap.
	We apply this margin regardless of whether an honest peer is resident in the target's routing table or simply part of the swarm.
	\item \vspace*{-0.5em}These two numbers are then added to arrive at the target score %
	that needs to be beaten by at least \texttt{lowWater} many of our malicious nodes in order to have the target's ConnMgr trim all connections to honest nodes (except for those within the grace period).
	\item \vspace*{-0.5em}In order to reach this score, we traverse the set of our malicious nodes and start relaying connections to randomly generated virtual nodes as required.
	We prioritise those of our nodes that are based on pre-generated identifiers, referred to as \emph{DHT nodes} to fill the target's routing table.
	If this is not sufficient (which is the case for higher-than-usual \texttt{lowWater} marks), we also boost the score of the random nodes that are run to reach \texttt{highWater} many connections to trigger the ConnMgr's disconnect routine.
\end{enumerate}

\vspace*{-0.5em}
Careful observation
 might suggest that \emph{precise} probing of the target's \texttt{lowWater} mark is required, since a too high estimate would cause those of our nodes that should occupy the lower buckets to be disconnected by the ConnMgr.
While this is technically correct, our nodes reconnect to the target within milliseconds.
This leaves only an extremely short window of opportunity for honest nodes to slip into the target's routing table.
This is due to the fact that routing table spots that become vacant during a disconnect wave are not automatically filled by remaining connections.
Instead, the \gls{dht} component of libp2p only reacts to a well-defined set of messages to insert peers into  a node's routing table.
One such message is a ping.
We exploit this behaviour and re-establish  any severed connection as soon as the ConnMgr executes its connection trimming routine and ping the target from all still connected nodes at the same time.
As we will elaborate in Section~\ref{sec:eval}, we have evaluated this strategy to be effective, as fully occupying any node's routing table takes mere minutes.
Moreover, completely eclipsing nodes with $ \approx75\% $ probability takes less than an hour.
In effect, this results in overall negligible costs, even when seeking to disrupt the complete public DHT-based IPFS network.

\subsection{Attack Wrap-Up}
Our attack exploits a variety of flaws in libp2p/IPFS, specifically in the uncoordinated coexistence of the \gls{dht} subsystem and the ConnMgr.
This results in nodes actively trimming connections to honest peers after mere minutes, due to the ease of tricking the ConnMgr.
As part of engaging in a responsible disclosure process with Protocol Labs, this has been acknowledged as a conceptual problem without an easy solution.
 The public DHT-based IPFS network, by design, accepts any node as long as it conforms to the protocol specifications.
Therefore, vulnerabilities that have been fixed in the reference implementation can be reintroduced into the network through nodes based on alternative implementations.

In summary, reaching a verdict regarding the global impact of our attack with respect to the whole IPFS ecosystem comprising the public DHT-based network, but also many other services based on IPFS technology (like DTube) requires careful investigation of each such system.

\definecolor{rt_attackers}{RGB}{255,0,0}
\definecolor{rt_others}{RGB}{0,179,0}
\definecolor{swarm_attackers}{RGB}{255,102,0}
\definecolor{swarm_others}{RGB}{0,255,0}
\definecolor{bucket_attackers}{RGB}{255,0,0}
\definecolor{bucket_others}{RGB}{0,179,0}
\definecolor{bucket_empty}{RGB}{0,0,179}

\section{Evaluation}
\label{sec:eval}
\begin{figure}[h!]
	\centering
	\subfloat[Default settings (\texttt{lowWater}=600, \texttt{highWater}=900, grace period=20s)\label{fig:swarm-rt-600}]{%
		
		\includegraphics[width=\columnwidth]{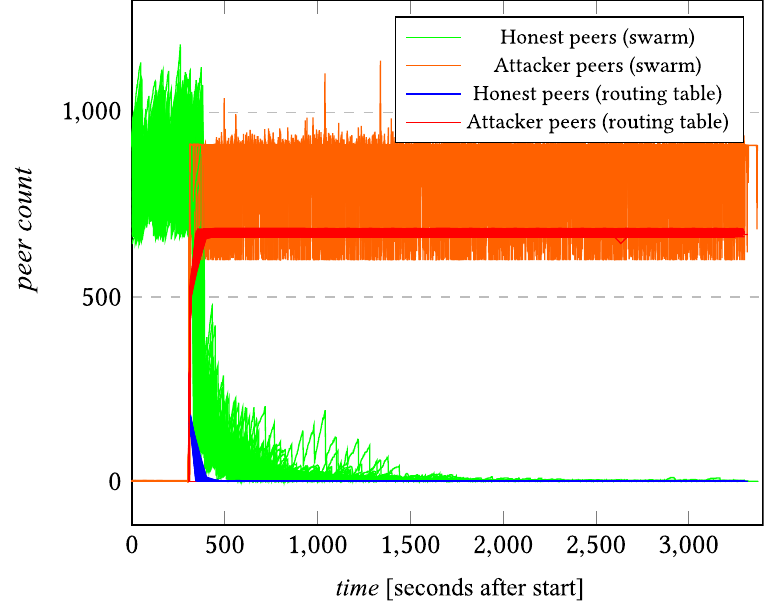}

}

\subfloat[Bootstrap settings (\texttt{lowWater}=1000, \texttt{highWater}=2000, grace period=60s). Two runs failed mid-way for reasons unknown. Note that only the routing table is relevant for an attack bootstrap nodes.\label{fig:swarm-rt-1000}]{%
	
		\includegraphics[width=\columnwidth]{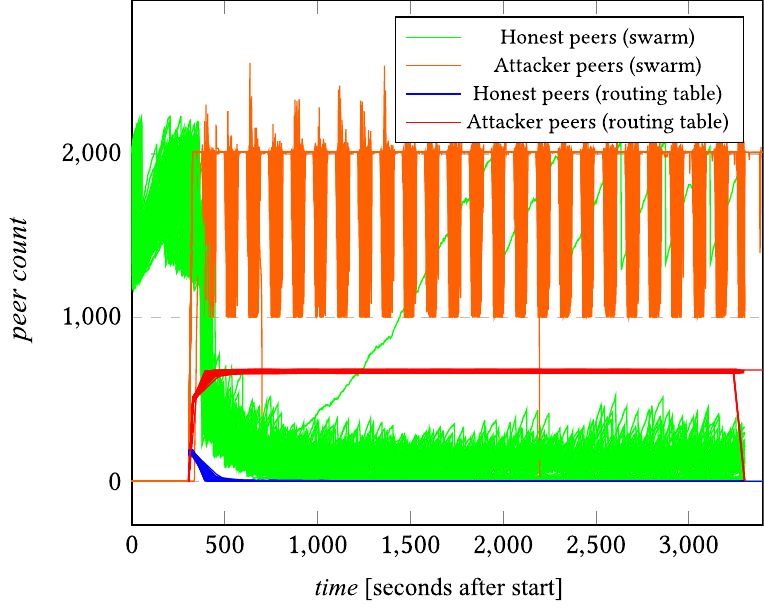}
			
}
	\caption{Visualisation of the number nodes in an attack target's swarm and routing table for 100 runs (overlaid). Regular operation is followed by a surge of malicious nodes after starting the attack.}
	
	\label{fig:rt-swarm-comparison}
\end{figure}

In order to gauge the impact of our attack, two key scenarios were evaluated.
Attack runs were carried out against unmodified IPFS nodes operated within the live IPFS network.
While our main attack target was go-ipfs 0.4.23, version 0.5.0 was also evaluated to asses the impact of countermeasures included in this release (see Section~\ref{sec:countermeasures}).
In both cases, the state of an attack target's swarm was queried locally at the target node, while routing table information was obtained remotely as part of the main attack loop (see Sections~\ref{sec:probing} and ~\ref{sec:atk:cont-probing}).
First, the impact on an average node with default parameters was measured.
Afterwards, nodes with the same configuration as the live IPFS bootstrap nodes\footnote{These parameters were kindly provided to us by Protocol Labs to help evaluating our attack's impact.} were attacked.
The evaluation results for both scenarios are provided in Figure~\ref{fig:rt-swarm-comparison}.
In both cases, the DHT is fully poisoned within minutes (Sections~\ref{sec:atk:defs} and~\ref{sec:atk:boos} provide in-depth discussions attack performance in each scenario).

Apart from attack performance evaluation, we provide an estimate regarding the cost of completely disrupting the whole live IPFS DHT for version 0.4.23.
The IPFS reference implementation, by default, combines Bitswap (which does not reach beyond immediate swarm connections) and DHT-based content routing to distribute data and locate other nodes.
The DHT component itself knows two modes of operation: \emph{client} mode, and \emph{full} mode  (also referred to as server mode).
Only those nodes presenting themselves as operating the full DHT, and advertising themselves as directly reachable on the IP network layer will be added to other nodes' local routing tables.
The set of all nodes meeting these requirements are critical to the public IPFS DHT.
Disrupting those nodes will also affect client-mode nodes, given that \enquote*{In IPFS, the DHT is used as the fundamental component of the content routing system}~\citep{schmahmannIPFSContentRoutinga}.
A collapsing DHT will have client-mode nodes rely purely on their immediate swarm connections for content discovery (see Section~\ref{sec:ipfs}).
As an immediate effect, the long tail of available data will not be available anymore.
Maintaining a disrupted state, however, will have severe effects:
Eclipsing all DHT server-mode peers means that all IP-layer information about honest nodes (which is ultimately used to connect to nodes) will vanish from the DHT.
Therefore, it is sufficient to take out server-mode nodes to also affect all clients.

The technical details on how this evaluation was carried out can be found in Appendix~\ref{appendix:eval}.
In general, every attack run was carried out 100 times against newly-spawned IPFS nodes with random IDs.

\subsection{Default Settings}
\label{sec:atk:defs}
The goal of this evaluation against an unmodified go-ipfs v0.4.23 node with default ConnMgr parameters (\texttt{lowWater} = 600, \texttt{highWater} = 900, 20s grace period) is to see whether we can eclipse average nodes in the IPFS network. This setting used an attack duration of 50 minutes.
Given that the pre-generated IDs amount to more than the default 600 \texttt{lowWater} many peers, poisoning a targets routing table and eclipsing the swarm is virtually equivalent. %
As a result, few relayed connections are sufficient to eclipse a target.

\begin{figure}[b!]
	\centering
	
		\includegraphics[width=\columnwidth]{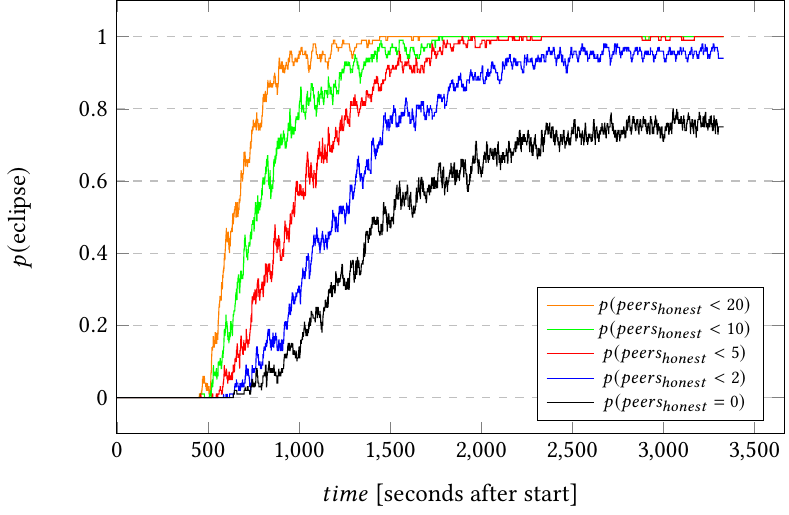}
	
\caption{Probability of eclipsing a node with default settings (\texttt{lowWater}=600, \texttt{highWater}=900, grace period=20s)}
\label{fig:eclipse-prob}
\end{figure}

Figure~\ref{fig:swarm-rt-600} visualizes the number of attackers and other nodes in the swarm and the target's routing table for all 100 runs.
First, the plot clearly shows that all 100 victims were well connected and that during the first 5 minutes the  \texttt{lowWater}  mark was never undershoot.
Starting the attack after 5 minutes, the number of attackers in the swarm and the routing table increase almost instantly, while the number of other nodes drops rapidly.
 After less than ten minutes, the routing tables of all attacked nodes are fully occupied by malicious nodes, while after less than 17 minutes, the probability of fully eclipsing a node is already $ >50\% $ as shown in Fig.~\ref{fig:eclipse-prob}.
In case an attacker's goal is not to fully eclipse a node but to prevent a node from discovering any content with high probability, even less time is required.
As also shown in Fig.~\ref{fig:eclipse-prob}, diminishing a target's swarm to less than ten connections is virtually guaranteed in less than half an hour.

\begin {figure}[!t]
\centering
\includegraphics[width=\columnwidth]{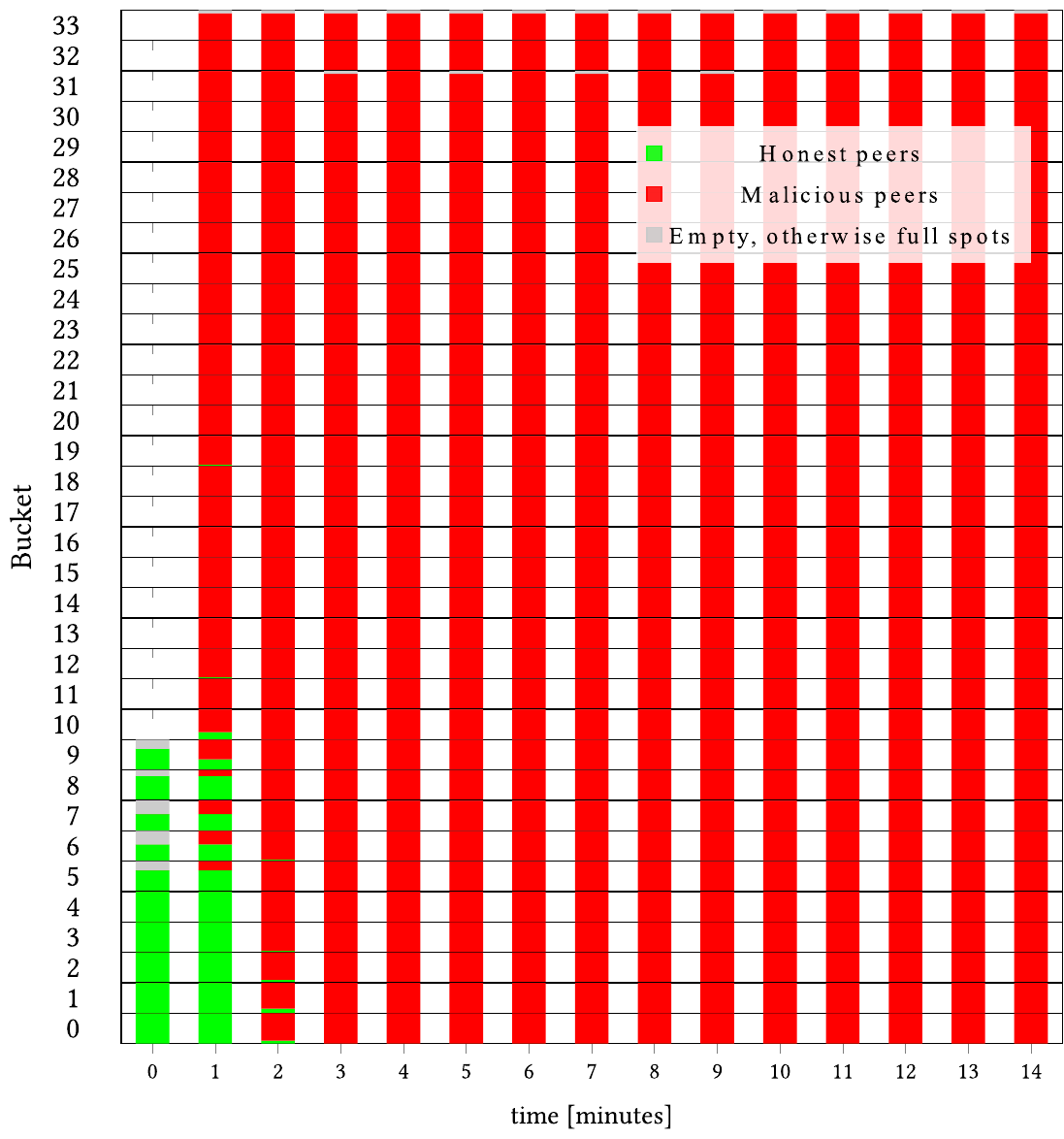}

\caption{Visualization of an attack target's routing table over the first 15 minutes for IPFS 0.4.23. Honest peers are green, malicious peers red, and empty spots in otherwise filled buckets are shown in grey.%
}

\label{fig:default_rt_buckets}
\end{figure}

To better illustrate the impact of our attack on a target's routing table, Figure~\ref{fig:default_rt_buckets} visualises a single target's routing table buckets.
This reveals that initially only about 9 buckets are full or partially filled, before the attack is started. 
The second time slot on the x-axis shows the state directly after the start of the attack.
All empty spaces in all buckets until bucket 33 are filled up by the attackers. 
The third time slot shows that after the connection manager tries to reduce the number of connections, the number of other peers decreases drastically, meaning the node has been tricked to harm itself.
After the next connection cleanup phase, the routing table is fully poisoned.%

\subsection{Bootstrap Settings}
\label{sec:atk:boos}
This setting uses an unmodified go-ipfs v0.4.23 node configured to the same settings as the official bootstrap nodes (\texttt{lowWater} = 1000, \texttt{highWater} = 2000, 60s grace period).
The goal of this evaluation is to see whether we can eclipse the official bootstrap nodes.
Attack duration was set to 50 minutes.
  
For bootstrap-node-like configurations, the number of peers using a pre-generated ID for routing table poisoning is not sufficient to have the ConnMgr prune all honest peers from an attack target's routing table.
As a result, the number of relayed connections can easily explode when seeking to fully eclipse such a node, which can result in inadvertently overburdening an attack target.
In order to keep this period of high strain as short as possible, we disable spawning relayed connections as soon as a target's \gls{dht} is fully poisoned.
As shown in Figure~\ref{fig:swarm-rt-1000} this does not result in a successful eclipse attack against an IPFS node run with bootstrap node configuration.
However, the swarm state is irrelevant for actual bootstrap nodes, as swarm connections are not used for bootstrapping.
In addition, as nodes are constantly joining the network, new swarm connections to the bootstrap nodes are established.
Therefore, attacking the swarm of bootstrap nodes is futile.
Still, even \enquote{only} poisoning the bootstrap nodes' routing tables results in all newly (re)connecting peers to only learn about malicious nodes, completely disrupting the live IPFS network over time.
Since this strategy is highly churn-dependent, the estimating time required to reach the whole network is out of this work's scope. %

\subsection{Attack Costs}
\label{sec:atk:costs}
In order to quantify the actual costs of running large-scale attacks against IPFS, key network metrics are required.
\citet{henningsenMappingInterplanetaryFilesystem2020} found on average 44474 concurrently online server nodes for the live IPFS DHT in early 2020.
Of these 44474 nodes on average only 6.55\% (about 3000 nodes) responded to connection attempts, meaning the remainder of nodes are likely operated by private individuals behind a NAT. 
As a consequence, it is sufficient to attack these 3000 nodes to prevent the network from responding to any queries for honest nodes connecting to the network.
In addition, permanently occupying the bootstrap nodes' routing tables will prevent any (re)connecting node from ever connecting to the live IPFS network.
Combining these two strategies will thus have a devastating effect.

\paragraph{Up-Front Costs}
The only up-front cost for our attack is related to ID generation (see Section~\ref{sec:atk:id}).
A quick search on a European price comparison service reveals a per-terabyte price of less than \officialeuro20 including VAT\footnote{\url{https://geizhals.eu/?cat=hdx&xf=5588_HDD}} for external hard drives.
This results in an up-front storage cost of less than \officialeuro600, to store 29TB of identifiers\footnote{Buying physical disks is significantly cheaper than current rates for cloud storage for the scope of our attack.}.
In order to actually generate this many identifiers, any commodity personal computer can be used, adding an estimated \officialeuro1000 for a machine dedicated to generating identifiers, featuring an eight-core \emph{AMD Ryzen 3700X} eight-core CPU\footnote{\textcolor{blue}{\ttfamily{}https://geizhals.eu/?cat=sysdiv\&xf=10863\_8\%7 E6764\_AMD\%7E6770\_Ryzen+3000}}.
This results in an estimated up-front cost of less than \officialeuro2000 including a large margin for electricity costs that easily covers the time it takes to generate the required amount of identifiers as outlined in Section~\ref{sec:atk:id}.

\paragraph{Running Costs}
The evaluation setup relied on virtual machines with 4 cores and 16GB RAM, which cost $\officialeuro0.031$ per hour and instance\footnote{\url{https://www.hetzner.com/cloud}} including VAT.
Mapping this to the 3000 reachable IPFS nodes discovered by
\citet{henningsenMappingInterplanetaryFilesystem2020} results in running costs of $3000 \times 0.031\officialeuro = 93\nicefrac{\officialeuro}{h}$ to attack all reachable nodes simultaneously.
This does not allow drawing conclusions regarding the overall number of active IPFS users, since it does not respect the network's churn rate.
However, this has no bearing on the cost of eclipsing all available nodes (as this number would not change, only individual attack targets would come and go).
Due to libp2p's routing table not keeping disconnected peers, a fully eclipsed node is known to be undiscoverable by the rest of the network.
Once this is achieved, the cost of keeping a node eclipsed drops significantly, since it is sufficient to maintain as little as four connections to keep the node from re-connecting to bootstrap nodes.
This scenario is no longer restricted by CPU or RAM utilisation.
However, any churn requires continuous runs of the full attack against newly joining nodes, which inflates costs.
As of IPFS 0.4.23, global attacks are therefore more economic when targeting bootstrap nodes.
\emph{In general, however, this scales linearly with network size/churn rate, enabling highly efficient, global attacks.} %

\paragraph{Overall Attack Costs}
Considering the 75\% success rate after running the attack for 50 minutes against nodes with default ConnMgr configurations, bootstrap nodes are an economic attack target depending on attacker budget and time constraints.
After all, once nodes are fully eclipsed, severing all connection to these nodes will simply trigger the bootstrap routine.
Alternatively, a global attacker can simply wait until nodes restart, given only the static set of preconfigured bootstrap nodes will be contacted.

Trying to keep bootstrap nodes eclipsed, does requires running the full attack, since newly joining nodes will continuously establish connections.
Still, at a cost of \officialeuro0.031 per hour for a cloud instance, attacking the eight bootstrap nodes used as of IPFS 0.4.23 results in running costs of $8 \times 0.031\officialeuro = 0.248\nicefrac{\officialeuro}{h}$.

\section{Countermeasures}
\label{sec:countermeasures}
Our attack only works because it is easily possible to mount Sybil attacks, which were first introduced by \citeauthor{douceurSybilAttack2002a} in \citeyear{douceurSybilAttack2002a}.
The easiest way to prevent our attack would hence be to prevent Sybil attacks.
This, however, is hardly feasible in practice given the open and decentralised-by-design nature of IPFS, and its key promise to let anyone participate in the network without central governance.
As \citeauthor{douceurSybilAttack2002a} put it: \enquote{With no logically central, trusted authority [$\ldots$] it is always possible [$\ldots$] to present more than one identity [$\ldots$]}~\citep{douceurSybilAttack2002a}.
This is further supported by a \citeyear{levineSurveySolutionsSybil2006} survey by \citeauthor{levineSurveySolutionsSybil2006} and by a more recent study by \citet{mohaisenSybilAttacksDefenses2013}.
Still, attackers can be severely hindered weaponising the \emph{operation} of large quantities of malicious network nodes.
The \emph{generation} of a large identifier set cannot practically be mitigated in decentralised systems when considering large-scale attacks.
After all, it took us four days to generate enough identifiers to target networks consisting of billions of nodes.
While \gls{pow}-based ID generation as proposed by \citet{baumgartKademliaPracticableApproach2007} would inflate up-front costs, \citet{prunsterHolisticApproachPeertoPeer2018} have reached the conclusion that this is is ultimately futile even for modern systems employing self-certifying identifiers and authenticated end-to-end encryption.
Even a moderately funded attacker could simply invest in enough computing power to counter any realistic \gls{pow} factor.

Presenting our attack to Protocol Labs as part of the responsible disclosure process has intensified an already ongoing effort of hardening IPFS/libp2p, resolving the issue of unconditional removal of \gls{dht} nodes (see Section~\ref{sec:issues:impl}) in the libp2p version that ships with go-ipfs v0.5, as released on April 28, 2020.
Apart from that, many other fixes were released, with go-ipfs 0.6 (published in June) effectively preventing casual attackers from carrying out the attack presented in this paper.

Protocol Labs allowed us to attack bootstrap nodes running go-ipfs 0.5 right after its release, as well as performing attack runs run on 0.6.
In accordance with the timeline of releases, major changes affecting our attack are highlighted for go-ipfs 0.5, followed by a discussion on 0.6 improvements as well as other fixes related the discovered vulnerabilities.

\paragraph{IPFS 0.5 Mitigations}
Amongst many other changes, the libp2p version shipping with go-ipf 0.5 introduced a \gls{dht} eviction strategy including periodic routing table refreshing and testing peer liveliness based on three parameters:
	(1) \emph{Time of last successful outbound query},
	(2) \emph{last time a peer was considered useful} (see below), and (3)
	 \emph{eviction grace period}, which depends on bucket size and refresh period; typically in the order of 45 minutes.
The first parameter is used during periodic routing table refreshes.
In case a peer has not been successfully queried within the grace period, a ping is issued.
Failure to respond results in eviction.
Consequently, stale peers are periodically removed from the \gls{dht} even if the buckets they reside in still feature vacant spots.
\\
A peer is considered useful, if, as part of a query it either responded first, or responding took less than twice the time of the fastest responding peer.
Whenever this condition is met, the last useful time is recorded.
In case a bucket is full, while another peer shall be added to this bucket, peers whose last useful time lies beyond the grace period are evicted.
However, responding to a query with an empty result, thus not being \emph{actually useful}, is also considered useful.
This prevents nodes from penalising honest peers that simply could not provide the data required to respond to a query, which helps new nodes join the network.
At the same time, however, becoming useful can be relatively cheap, since no information is required to still become useful.
In addition, only routing table peers are evaluated for usefulness, due to the usefulness definition being limited to \gls{dht} operations.
Thus, even a theoretic swarm node providing all content ever queried would not join the routing table easier than any other node.

\begin {figure}%
\centering

\subfloat[Attack performance on a IPFS 0.5 node run with bootstrap node configuration over twelve hours. This node baehaved like a regular node and was not used for actually bootstrapping. Thus, it more closely resembles a regular node, not a bootstrap node with respect to network interactions.\label{fig:default_rt_buckets_05}]{%
\begin{adjustbox}{width=1\columnwidth}

	\includegraphics[width=\textwidth]{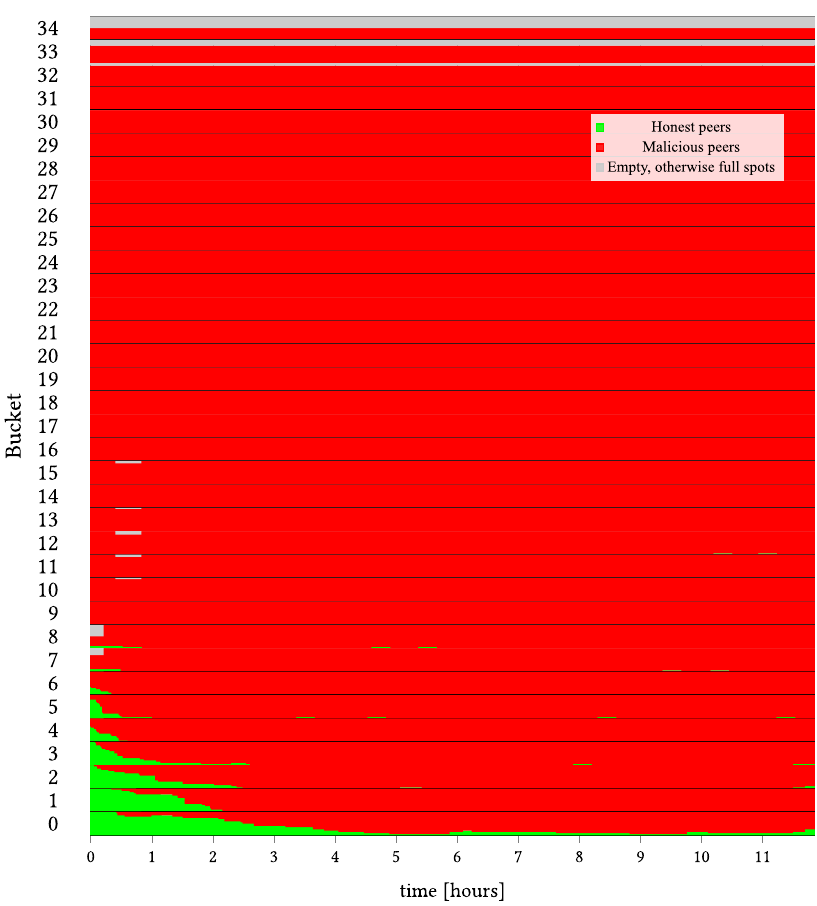}

\end{adjustbox}
}

\subfloat[Attack performance on a live IPFS 0.5 bootstrap node over three hours.  The attacks was launched 15 minutes after starting to collect metrics.\label{fig:bootstrap_rt_buckets_05}]{%
\includegraphics[width=\columnwidth]{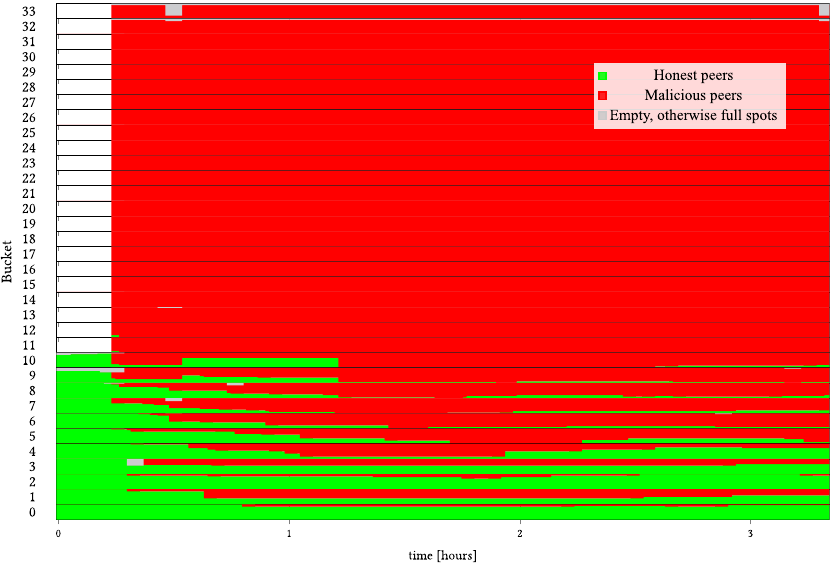}	
}

\caption{Visualisation of routing tables for go-ipfs 0.5. Higher attack resiliency can be observed for the bootstrap node due to interacting with more nodes.
	In comparison, the attack on a regular nodes grows increasingly successful over the first five hours. Honest peers are visualised in green, malicious peers in red and empty spots in otherwise filled buckets are shown in grey.%
}
\label{fig:rt_buckets_05}
\end{figure}

As a consequence of these changes, the connection trimming of the ConnMgr has no immediate effect on the routing table, since even disconnected peers are remembered and are reconnected to if necessary.
Thus, \texttt{lowWater}, \texttt{highWater} marks, and grace period also have no effect on the routing table.
Given that the typical eviction grace period is close to an hour, the cost of poisoning a target's routing table is increased several orders of magnitude compared to mere minutes to fully take over an IPFS 0.4.23 node's routing table.
In fact, even after twelve hours, an IPFS 0.5 node in bootstrap configuration still features honest peers in its routing table (see Fig.~\ref{fig:default_rt_buckets_05}).

Shortly after releasing go-ipfs 0.5 on April 28, 2020, Protocol Labs let us evaluate our attack against one of the production IPFS bootstrap nodes over three hours in the live IPFS network in order to gauge the impact of the  newly deployed countermeasures.
Since this concrete attack run targeted a production system, the evaluation period was chosen such that results could be obtained to make an initial judgement regarding the deployed countermeasures' utility without causing any real disruption to the network.
The results are presented in presented Fig.~\ref{fig:bootstrap_rt_buckets_05}.
Compared to attacking nodes specifically for evaluation purposes (not used by others for actual bootstrapping), an increased resilience can be observed.
This was to be expected, since the increased query frequency to bootstrap nodes awards a larger set of honest peers a \emph{useful} state.
As a consequence, these peers are kept in the routing table.
However, after only one hour, the initial number of 200+ honest peers in the bootstrap node's routing table could be more than halved and stayed below 100, clearly demonstrating the impact our attack still has on IPFS 0.5.
Protocol Labs's estimate regarding attack difficulty for 0.5, is that fully poisoning a bootstrap node's routing table is still possible within several days.

\paragraph{IPFS 0.6 and 0.7 Measures}
The most crucial measure to boost attack costs by orders of magnitude was included in go-ipfs 0.6 released in June 2020.
In effect, IP-addresses are now considered with respect to adding nodes to the local routing table, making it impossible to mount large-scale Sybil attacks from a single host.
Instead, at most three nodes associated with a single host (IPv4, IPv6\footnote{Even presenting multiple IPv6 addresses from a single large IPV6 subnet has no effect.}) can become resident in a node's routing table.
Therefore, this measure by itself already inflates the cost of attacking a single node by over two orders of magnitude compared to IPFS 0.5.
Mapping this against the estimate of several days to fully eclipse a bootstrap node clearly puts this out of reach for casual attackers.
Compared to theoretic proposals of limiting the numbers of identifiers that can be advertised to a \gls{p2p} network per IP address, multiple nodes can still join the network from the same host.
In fact, this aligns with the distinction between client and server-mode DHT that aims at mapping to nodes operated behind \glspl{nat}.
In addition, IPFS 0.7 deprecated the previously used transport security mechanism, which breaks compatibility to pre-0.6 releases.
As a result, older instances are required to update to a release containing fixes, if operators want to continue to participate in the network.

\paragraph{Additional Deployed Changes}
Aside from these key changes, additional countermeasures were rolled out since IPFS 0.4.23.
These include a fixed score for nodes acting as relays to prevent relay-based inflation of ConnMgr scores.
While many third-party projects rely on different parts of libp2p's functionality, the relaying subsystem is expected to be used in virtually all \gls{p2p}-related projects that require connecting users behind \glspl{nat}.
Therefore, this change is expected to impact many libp2p-based applications.
Moreover, nodes in the lowest two \gls{dht} buckets are exempt from pruning, while higher buckets are now assigned a fixed score of five points.
While this may seem counter-intuitive, it prevents attackers capable of generating a huge body of valid identifiers from gaining an advantage for small network sizes by ConnMgr points awarded from the \gls{dht} subsystem.
In effect, nodes are scored more equally due to these changes.
Moreover, libp2p now verifies reachability of advertised IP addresses, and only adds those nodes who actually respond to connection requests.
This already goes a long way towards solving the issue of unconditional trust and makes it harder to become resident in routing tables.
In addition, configuration of direct peering agreements was also elevated to a core feature of the IPFS reference implementation.

\section{Related Work}
\label{sec:related}
Our work presents an attack on IPFS, a decentralised peer-to-peer system in production use.
Therefore, this section mainly focuses on work related to analysing and attacking similar live systems, since theoretical models and surveys regarding countermeasures to eclipse and Sybil attacks have been extensively published before.
For these topics, we refer to the works of \citet{levineSurveySolutionsSybil2006} and \citet{mohaisenSybilAttacksDefenses2013} already discussed in Section~\ref{sec:countermeasures}.

Although IPFS was launched in 2013 and has been continuously gaining traction, scientific literature on IPFS is scarce.
Apart from the original whitepaper describing its system properties~\citep{benetIPFSContentAddressed2014} and protocol specifications of varying maturity\footnote{\url{https://github.com/ipfs/specs}}, little in-depth documentation on IPFS is currently maintained.
Metrics on the overall IPFS network are also not available.
However, a \citeyear{henningsenMappingInterplanetaryFilesystem2020} paper~\citep{henningsenMappingInterplanetaryFilesystem2020} crawled the live IPFS network, whose results were used to estimate the cost of our attack in Section~\ref{sec:eval}.
This work also stressed that no countermeasures against Sybil attacks were in place as of version 0.4.23, contrary to the original IPFS whitepaper.
The authors also observed that the way IPFS implements content discovery (first querying all connected peers, only then falling back to querying the DHT in case none of the contacted peers would serve the requested content) provides some resilience against eclipse attacks aimed at the DHT.
As our work demonstrates, however, this defence was extremely limited, making it still possible to fully eclipse a node in less than one hour with $ \approx75\% $ probability.

Similar to our work on IPFS, \citet{heilmanEclipseAttacksBitcoin2015} demonstrated a successful eclipse attack against \emph{Bitcoin}~\citep{nakamotoBitcoinPeertoPeerElectronic2008}.
On the surface, this attack was based on flaws similar to the issues we discovered in IPFS.
Most prominently, peers that were still running and maintaining connections to an attack target could be replaced by fresh malicious peers.
Of the countermeasures deployed to the Bitcoin reference implementation, an adapted eviction strategy---keeping live peers instead of having them easily replaced---showed the most impact towards defending eclipse attacks.
Actually carrying out an eclipse attack against even a single Bitcoin node without such countermeasures is considerably more expensive than our attack against  prior to version 0.6, due to the fact that widely-distributed IP addresses are required.
This was mainly due to the different organisation of Bitcoin's routing tables compared to the libp2p \gls{dht} and not primarily for reasons of hardening against attacks.
\\
An attack that uses similar principles than ours has been shown to be successful against the \emph{Ethereum} cryptocurrency~\citep{marcusLowResourceEclipseAttacks}\footnote{This attack was performed prior to Ehtereum's planned switch to libp2p.}.
However, there are several important differences to our work.
Most prominently, our attack is more powerful, as it works in near-realtime by tricking a node into actively disconnecting itself from the rest of the network.
In contrast to the attack on Ethereum, our attack thus requires no other forms of \gls{dos} attacks to succeed.
This is crucial, as system downtime could be easily detected, while our attack simply requires establishing additional connections (many of which can be multiplexed, thus not showing up on network monitors).
The general issue of how IPFS implements connection management is distinctly different from Ethereum and also independent of the DHT-related attack vector that shows similarities to the attack on Ethereum.
As such, our attack is not limited to the DHT subsystem that shares much of the conceptual functionality of Ethereum's \gls{p2p} network.
Moreover, our way of gaming IPFS's ConnMgr makes it possible to deliberately have nodes outside an attackers control induce churn with low effort.
This in itself is an attack not discussed in related works targeting systems used in practice\footnote{Evaluating the impact of such churn attacks is out of this work's scope}.
Our attack against IPFS v0.4.23 also takes only minutes to fully poison any node's routing table and less than an hour to fully eclipse a node with high probability, without requiring additional \gls{dos} attacks.
In addition, countermeasures such as non-public mapping from node id to bucket are simply not applicable to IPFS as it would render one of it core features defunct.
In general, IPFS is a more complex overall system and the conceptual issue of connection management persists. As of IPFS 0.6.1, this general issue persists, as it is inherently difficult to tackle, although no cheap attacks are currently known.
In summary, the attacks against cryptocurrencies require the target to reboot, take several days to execute, are orders of magnitude more expensive, and suggested countermeasures are not applicable to IPFS.
The same applies to the follow-up attack on Ethereum by \citet{henningsenEclipsingEthereumPeers2019}.\\
Most importantly, however, previous estimates for global attacks provided in these related works do not consider larger network sizes, which would exponentially inflate the cost of identifier generation.
Our work, in contrast, scales linearly with a network's size as identifier generation is a constant factor, even accounting for networks of billions of nodes.
In addition, with Ethereum's switch to libp2p, our findings can serve as basis for further, in-depths analysis and hardening efforts that also benefit the Ethereum community.

In the context of content distribution, a survey of \emph{BitTorrent}~\citep{cohenBitTorrentProtocolSpecification2013} is of significance, since IPFS's Bitswap is based on the way BitTorrent distributes content among peers.
Carried out in \citeyear{piatekIncentivesBuildRobustness2007} when BitTorrent was immensely popular, the work by \citet{piatekIncentivesBuildRobustness2007} revealed inherent flaws in the way peers are awarded for providing content to others.
One of this work's conclusions is that a small minority of altruistic peers are responsible for BitTorrent robust performance in real-world settings.
This is understandable given that ways to incentivise users to provide resources to others without any certainty of receiving something in return is inherently difficult, especially in low-trust environments.
Protocol Labs has recognised this launched the \emph{Filecoin} cryptocurrency in summer 2020 as an incentivisation layer atop IPFS. %

\section{Conclusions}
\label{sec:conclusions}

In this paper, we introduced, described, and demonstrated a successful end-to-end eclipse attack on IPFS.
Exploiting vulnerabilities of the libp2p library, one aspect of our attack is to successfully poison routing information locally stored by nodes of IPFS's underlying P2P network. 
In addition, we have shown that our attack enables us to eclipse arbitrary IPFS nodes and, consequently, to disrupt the entire public DHT-based IPFS network step-by-step.
Applying our attack on live IPFS nodes demonstrated the attack's effectiveness and feasibility.
Alarmingly, conducted evaluations have shown that even global attacks can be already mounted with moderate efforts, making this attack an option also for attackers with limited resources.
\emph{As such, we have shown that our global attack scales well with a network's size, fully answering our initial research question.}

The impact of our proposed attack is substantial mainly for two reasons.
First, our attack exploits a conceptual flaw in the connection management of IPFS with no easy solution.
Secondly, our work has lead to a successful, ongoing hardening process.
The whole ecosystem beyond decentralised exchange of data (but also other IPFS-based services) benefit from upstream releases incorporating fixes.
Two major version of the IPFS reference implementation were released since roprting our findings to Protocol Labs, both of which contained fixes, resulting in a huge increase of attack costs.
As a consequence, casual attackers won't be able to replicate our attack against updated nodes.

\printbibliography
\appendix
\begin{figure*}[h!]
	\centering\includegraphics[width=0.8\textwidth]{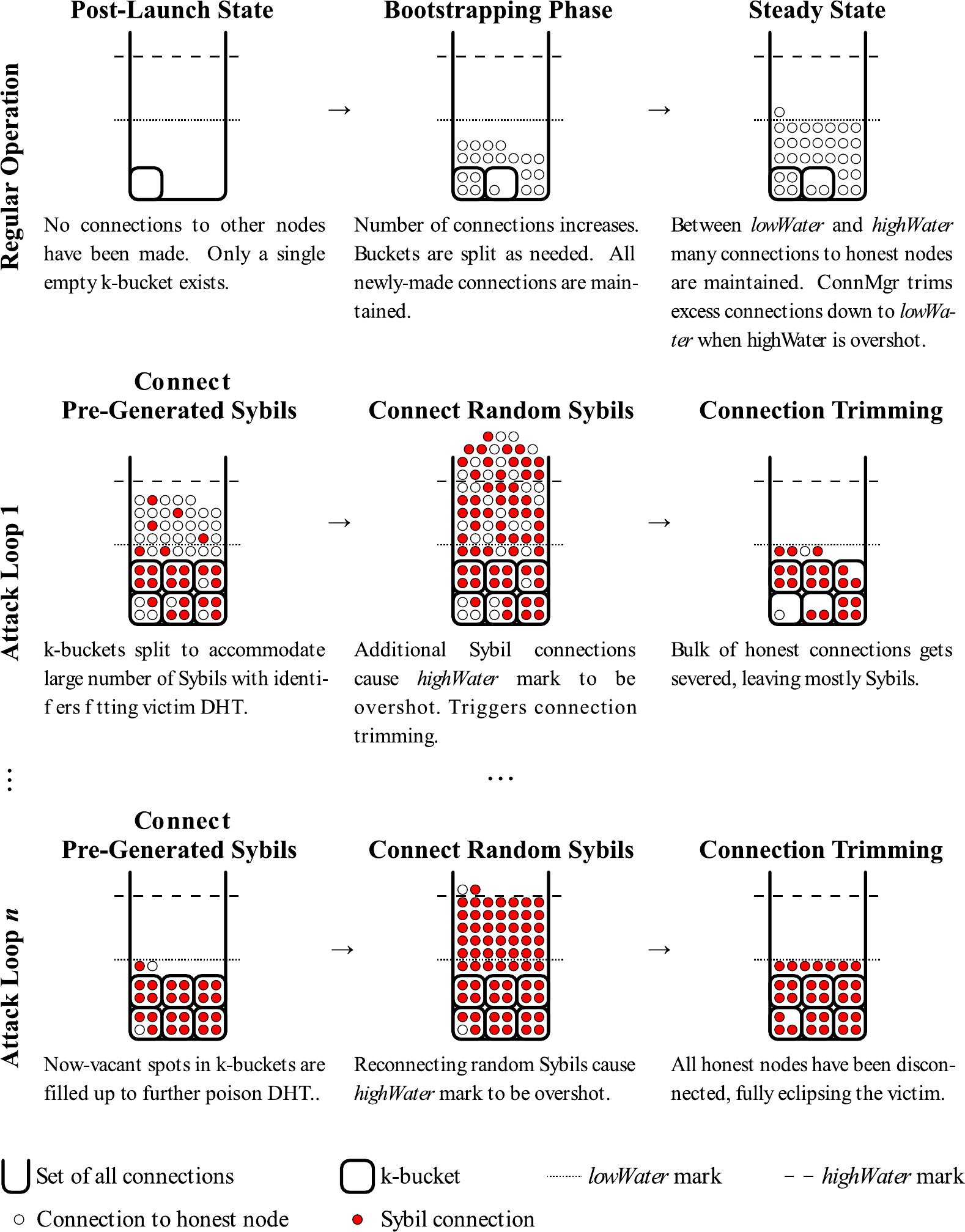}
	\caption{Schematic view of an attack victim’s connections and k-buckets (DHT) lifecycle spanning node launch, bootstrapping, steady-state regular node operation and total eclipse.}
	\label{fig:bucketprogression}
\end{figure*}

\section{Additional Attack Visualisations}
\label{app:vis}
Fig.~\ref{fig:bucketprogression} presents a high-level illustration of the progression of a victim node's swarm (including buckets) during normal operation, as well as during our attack. It has to be noted that the connections are depicted as an unordered set (with buckets being unordered subsets).
Therefore, trimming is not ordered top-to-bottom in this visualisation.
This choice was made to reflect two real-world aspects of the connection trimming routine:
Connection's scores do not necessarily correlate with their position within the buckets or even throughout the ordered set of buckets as it is implemented by the DHT subsystem.
Moreover, connections' scores change overtime, causing them to be reordered between invocations of the connection trimming routine.
Therefore, depicting connections as unordered comes closer to reality.

\clearpage

\section{Ethical Considerations}
\label{sec:ethics}
Mounting attacks on existing solutions used in practice and publishing details on these attacks raises ethical issues.
This especially applies to the work presented in this paper, as we have mounted our attack also on the live network for evaluation purposes.
Although all attacks have been run in the scope of a responsible-disclosure process and have been closely aligned and coordinated with Protocol Labs, we are aware that this kind of research needs to be subject to ethical considerations, which we hence discuss in this section.

As potential security impacts of the found vulnerabilities were apparent right after its discovery in April 2020, \textbf{we immediately contacted {Protocol Labs}}, initiating a responsible disclosure process.
In the scope of this process, the vulnerability that serves as basis for our attack has been assigned CVE-2020-10937\footnote{\url{http://cve.mitre.org/cgi-bin/cvename.cgi?name=2020-10937}}.
Furthermore, we have closely aligned follow-up activities with Protocol Labs to prevent negative impacts on the IPFS live network while conducting further research.
Protocol Labs has approved submitting this paper and even provided multiple rounds of feedback and helped ensuring factual accuracy.

Negative impacts have been prevented by the following measures:
\begin{itemize}
\item In general, attacks have been run on self-operated IPFS nodes only to avoid negative effects on third-party nodes.
\item By attacking a single self-operated node at a time, connectivity to this node was impaired, rendering only this specific node invisible to the rest of the network.
This has no practical side effects, due to the inherent redundancy of the network.
\item When evaluating our attack on bootstrap nodes run by {Protocol Labs} (see Section~\ref{sec:eval}), only one was been attacked.
However, four out of eight nodes in total are used for bootstrapping as of IPFS 0.5.
Hence, running our attack on one node provided tangible results without causing adverse effects.
\end{itemize}

Closely involving Protocol Labs and carefully coordinating each step of our attacks assured that our research did not harm live deployments.
Protocol Labs actively supported our research, e.g. by monitoring bootstrap nodes under attack and providing direct monitoring of nodes under attack.
Overall, the team at Protocol Labs acted professionally, actively supported us in evaluating our attack against core IPFS infrastructure, and invited further research based on our discoveries.
Public disclosure of the found vulnerability has been coordinated with Protocol Labs for October 2020 which includes publishing this eprint version of this paper and a blog post on the official IPFS blog\footnote{\url{https://blog.ipfs.io/2020-10-30-dht-hardening/}}.
At this point in time, a hardened version of IPFS will have been available for several months, which includes mitigation and substantially raises the bar for exploiting the vulnerability described in this work.
The work presented in this paper contributed to the development of this hardened version of IPFS.

\section{Evaluation Procedure}
\label{appendix:eval} 
The generic attack setup used for evaluating  attack performance consisted of  
one control server and one ID server (see Section~\ref{sec:atk:id}), as well as one virtual machine acting as victim and one acting as attacker for each evaluation run.
The victim and the attacker have been deployed at different hosters and regions. 
The control server can control an arbitrary number of victims and attackers simultaneously.
\emph{nmon}\footnote{\url{http://nmon.sourceforge.net/}} was used to collect CPU, memory, and network metrics both for victim and attacker instances.
A so-called \emph{swarm-monitor} script was used at the victim to monitor swarm state (attackers vs. honest peers).
The main attack binary is referred to as \emph{ipfs\_atk}.

For every evaluation/run the following procedure was carried out:
\begin{enumerate}
	\item Clean up the target, by killing all \emph{nmon}, swarm-monitor, and \emph{ipfs} processes and deleting the IPFS configuration. 
	\item Killing  all \emph{nmon} and \emph{ipfs\_atk} processes on the attacker machine.
	\item Initialise the IPFS configuration.
	 Additionally, the  \texttt{lowWater}, \texttt{highWater} marks, and the \emph{grace period} values are set depending on the evaluation scenario.
	\item Start the IPFS daemon on the target machine.
	\item Query the \emph{ID} of the target through the main IPFS network.
	\item Prepare the attacker by loading the pre-generated IDs for the attacker nodes from the ID server and generating all files necessary for the attack. 
	\item Query the target periodically for the number of connected peers. This step is repeated until the number of peers surpasses the configured \texttt{highWater} value, to ensure the target is well connected.
	\item After this threshold is reached, the logging scripts and processes are started. 
	\item Next, wait for 5 minutes and start the attack afterwards.
	\item Let the attack continue for a configurable time-period.
	This period is referred to as \emph{attack duration}.
	\item On the target, the number of peers, grouped in two categories, attacker nodes and honest nodes are recorded by the \emph{swarm-monitor} script.
	On the attacker side, the results of periodic routing table poisoning are logged.
	On both machines CPU, RAM and network utilisation is logged by \emph{nmon}.
	\item After the attack duration period all logging-related script and processes on the target and the attacker machine are stopped, followed by the attack itself.
	\item After each run, logs are downloaded and the target machine is rebooted to prevent impacts of previous attacks on future runs. 
	\end{enumerate}
\newpage
\clearpage
\section{Scaled-Up Evaluation Figures}
This appendix provides scaled-up versions of the original figures found in the main paper for a more legible, detailed interpretation of results.
\vfill\null

\begin {figure}[h]
\centering
	
	\includegraphics[width=\columnwidth]{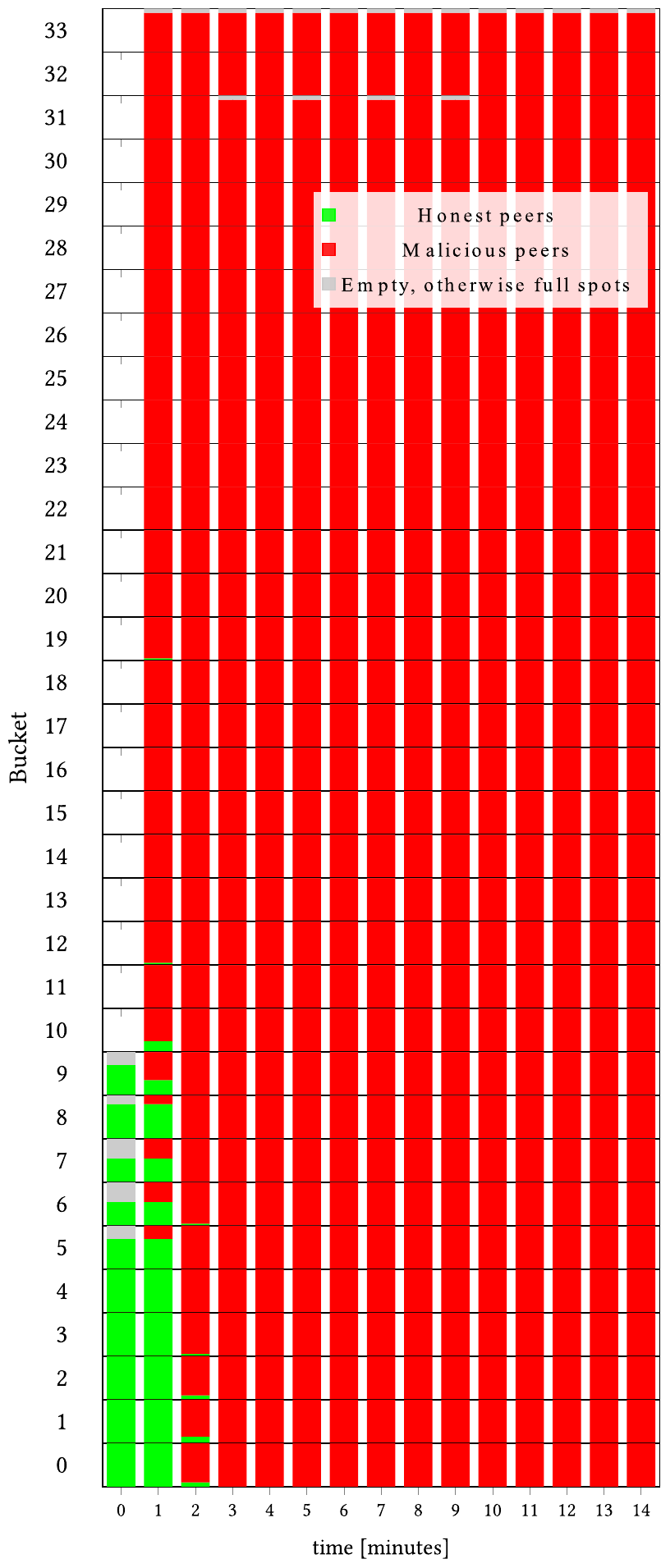}
\caption{Re-scaled version for Figure~\ref{fig:default_rt_buckets}: Visualisation of an attack target's routing table over the first 15 minutes for IPFS 0.4.23. Honest peers are visualized in green, malicious peers in red and empty spots in otherwise filled buckets are shown in grey. The x-axis shows time slots one minute apart and the y-axis shows buckets.}
\label{fig:default_rt_buckets_scaled}
\end{figure}

\begin{figure*}
	\centering
	\subfloat[Default settings (\texttt{lowWater}=600, \texttt{highWater}=900, grace period=20s)\label{fig:swarm-rt-600_scaled}]{%
		
		\includegraphics[width=0.6\textwidth]{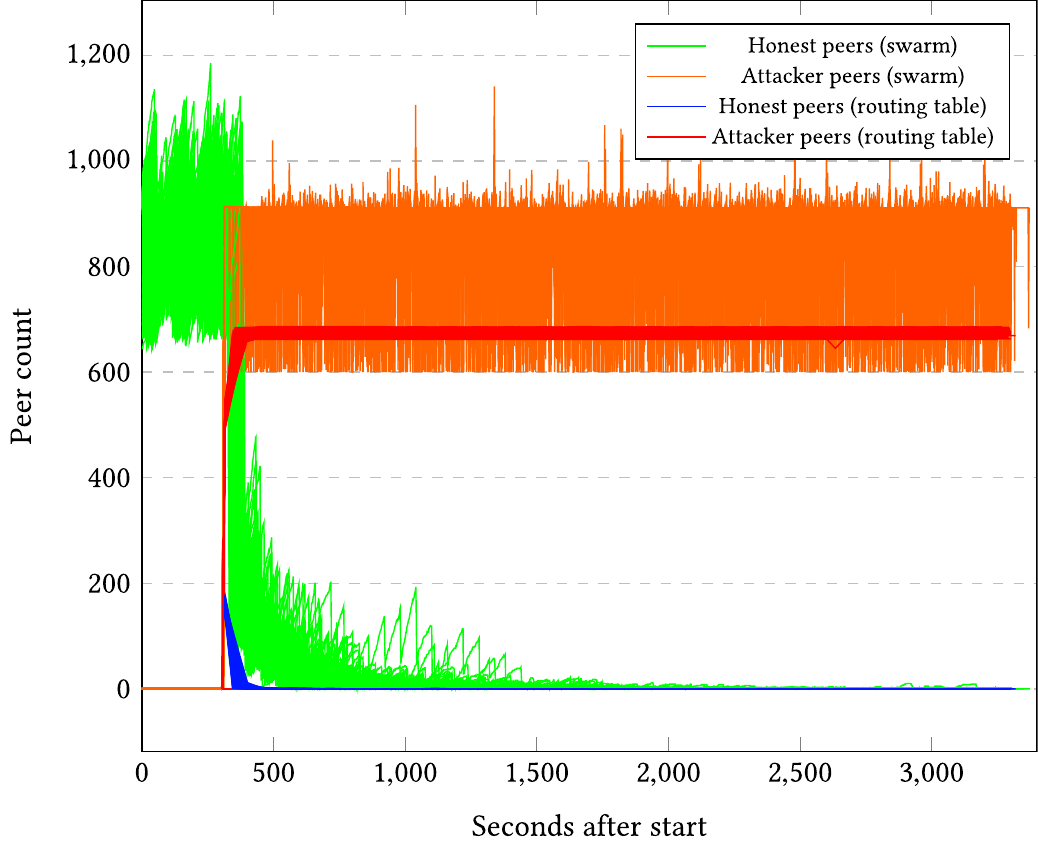}
		
}

\subfloat[Bootstrap settings (\texttt{lowWater}=1000, \texttt{highWater}=2000, grace period=60s). Two runs failed mid-way for reasons unknown.\label{fig:swarm-rt-1000_scaled}]{%
	
	\includegraphics[width=0.6\textwidth]{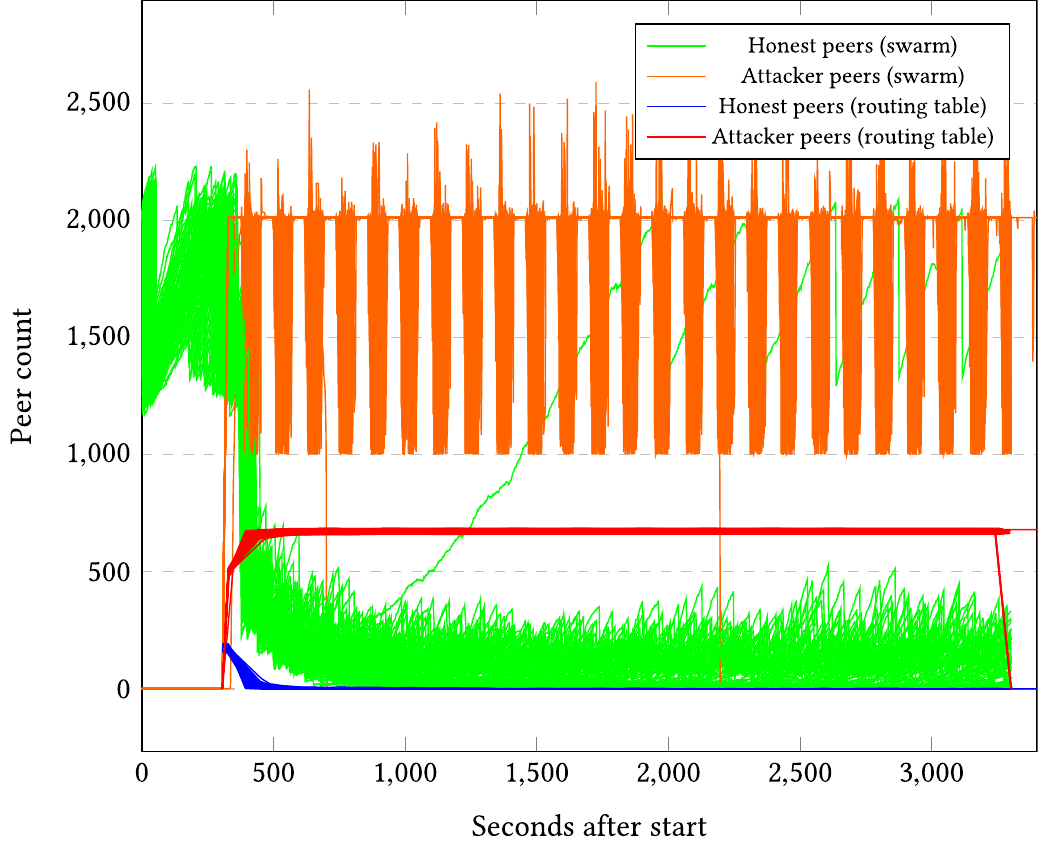}
}
	\caption{Re-scaled version of Figure~\ref{fig:rt-swarm-comparison}: Visualisation of the number nodes in an attack target's swarm and routing table for 100 runs. The red/orange graphs show the number of attacker nodes, while the green/blue ones show the number of honest nodes.}
	
	\label{fig:rt-swarm-comparison_scaled}
\end{figure*}
\clearpage

\begin{figure*}
	\centering
		\includegraphics[width=\textwidth]{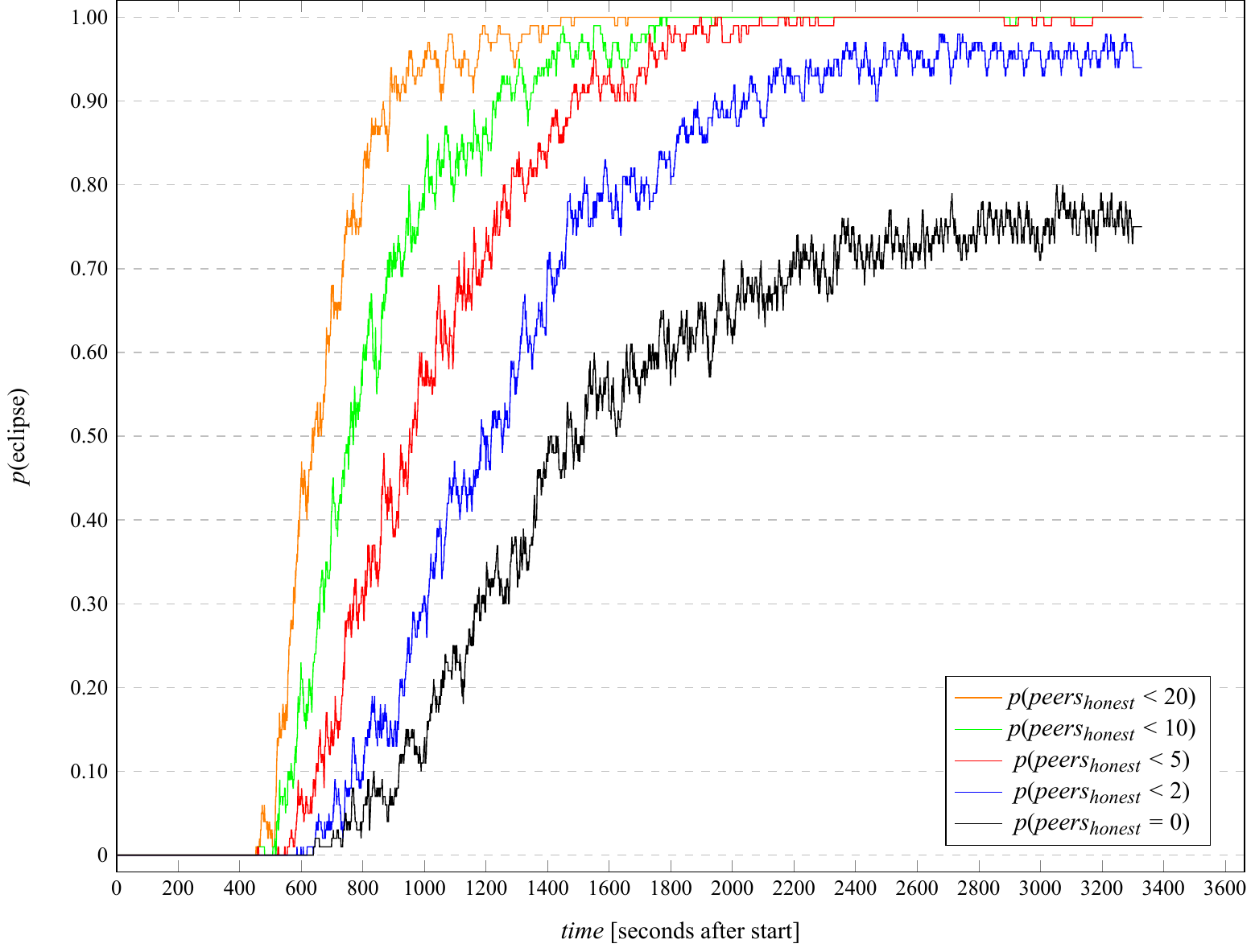}
\caption{Re-scaled version of Figure~\ref{fig:eclipse-prob}: Probability of eclipsing a node with default settings (\texttt{lowWater}=600, \texttt{highWater}=900, grace period=20s)}
\label{fig:eclipse-prob_scaled}
\end{figure*}

\begin {figure*}%
\centering

\includegraphics[width=\textwidth]{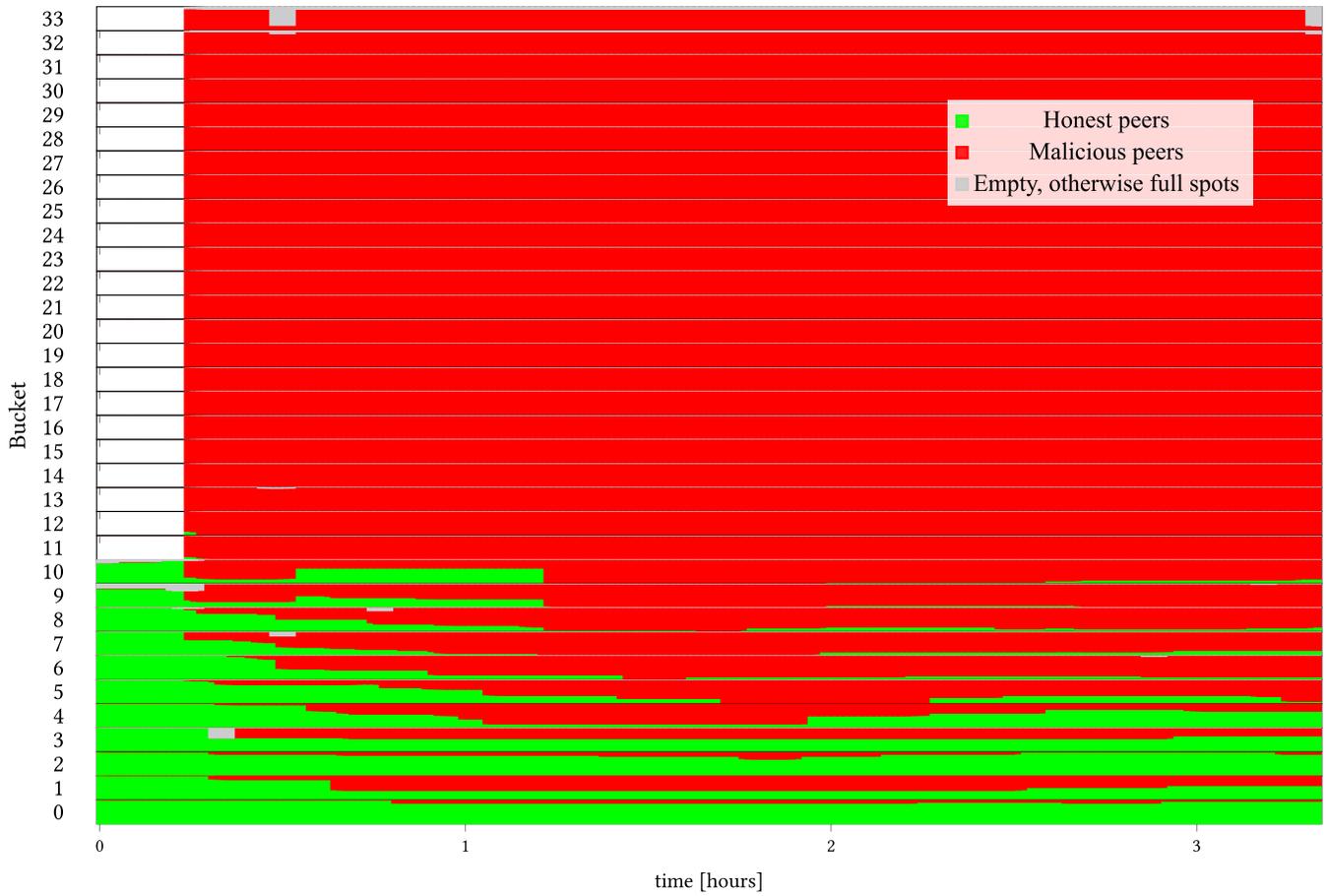}

\caption{Re-scaled version of Figure~\ref{fig:bootstrap_rt_buckets_05}: Visualisation of the bootstrap node's routing table over an attack duration of three hours for IPFS 0.5.0. Honest peers are visualised in green, malicious peers in red and empty spots in otherwise filled buckets are shown in grey. The x-axis shows time slots one hour apart and the y-axis shows buckets. The attacks was launched 15 minutes after starting to collect metrics.}
\label{fig:xxx}
\end{figure*}

\end{document}